\documentclass[acmsmall]{acmart}

\setcopyright{acmlicensed}
\acmJournal{PACMHCI}
\acmYear{2023} \acmVolume{7} \acmNumber{CSCW1} \acmArticle{121} \acmMonth{4} \acmPrice{15.00}\acmDOI{10.1145/3579597}

\usepackage{caption}
\usepackage{booktabs}
\usepackage{subcaption}
\usepackage{graphicx}
\usepackage[export]{adjustbox}
\AtBeginDocument{%
  \providecommand\BibTeX{{%
    \normalfont B\kern-0.5em{\scshape i\kern-0.25em b}\kern-0.8em\TeX}}}

\usepackage{caption}
\usepackage{subcaption}
\usepackage{enumitem}

\newcommand{\zoomviews}{
\begin{figure}
\centering
\begin{subfigure}[t]{.4\textwidth}
  \centering
  \includegraphics[width=.9\linewidth]{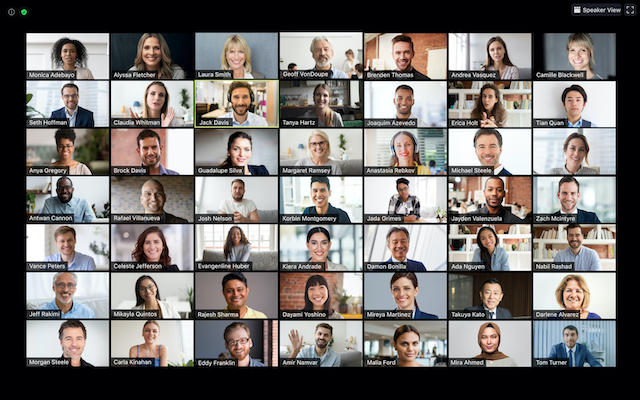}
  \subcaption{Gallery View}
\end{subfigure}
\begin{subfigure}[t]{.4\textwidth}
  \centering
  \includegraphics[width=.9\linewidth]{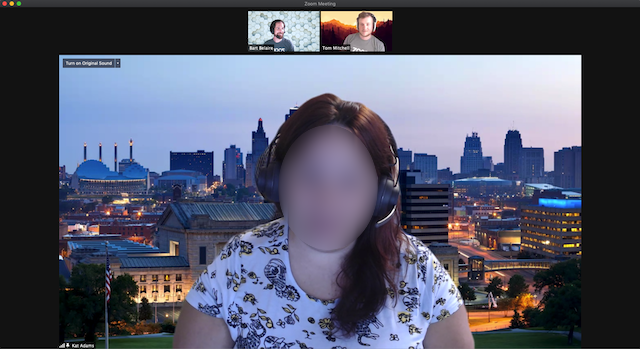}
  \subcaption{Speaker View}
\end{subfigure}
\caption{The two primary user views on Zoom. Images retrieved from Zoom's support website on May 25, 2022: https://support.zoom.us/hc/en-us/articles/201362323-Adjusting-your-video-layout-during-a-virtual-meeting}
\label{fig:zoomviews}
\end{figure}
}

\newcommand{\demographics}{
\begin{table}[]
\centering
\small
\begin{tabular}{@{\hspace{0cm}}lllll@{}}
\hline
\textbf{ID} & \textbf{Gender} & \textbf{Race and Ethnicity} & \textbf{Age} & \textbf{Job Level} \\ \hline
P1 & Male &  & 25-29 & Intermediate level \\
P2 & Male & White & 45-49 & Executive-level management \\
P3 & Male & White & 50-54 & Middle-level management \\
P4 & Male & Filipino & 25-29 & Intermediate level \\
P5 & Female & Asian (Korean) & 25-29 & First-level management \\
P6 & Male & Asian & 35-39 & Intermediate level \\
P7 & Female & White & 25-29 & Intermediate level \\
P8 & Female & White & 45-49 & First-level management \\
P9 & Female & White & 35-39 & Intermediate level \\
P10 & Female & Asian & 20-24 & Entry level \\
P11 & Female & White (non-Hispanic) & 20-24 & Entry level \\
P12 & Male & White & 45-49 & Middle-level management \\
P13 & Non-binary transmasculine & White & 30-34 & Intermediate level \\
P14 & Female & White & 40-44 & Middle-level management \\
P15 & Female & White & 20-24 & Entry level \\
P16 & Female & Korean-American & 25-29 & Entry level \\
P17 & Female & White Irish & 45-49 & First-level management \\
P18 & Female &  & 30-34 & Middle-level management \\
P19 & Male & Asian & 20-24 & Entry level \\
P20 & Female & Black & 30-34 & Middle-level management \\
P21 & Male & Black and white, non-Hispanic & 45-49 & Middle-level management \\
P22 & Male & Mixed Hispanic & 25-29 & Intermediate level
\end{tabular}
\caption{User demographics. All questions were optional, and all but job level were free-response. Some responses under ``Race and Ethnicity'' were modified for brevity, but original granularity was maintained. Blank cells indicate no response.}
\label{tab:demographics}
\end{table}
}

\newcommand{\summarytable}{
\begin{table}[]
\renewcommand{\arraystretch}{2}
\small
\setlist[itemize]{leftmargin=*}
\centering
\begin{tabular}{@{\hspace{0cm}} p{0.22\linewidth} p{0.25\linewidth}  p{0.25\linewidth}  p{0.25\linewidth} @{}}
\hline
\textbf{Feature} & \textbf{Effects on Bias} & \textbf{Causes of Leader Control} & \textbf{Recommendations} \\ \hline
\textbf{User Tiles} & User tiles prioritize displaying participants whose cameras are on, making them even more likely to participate and pushing others to the ``back row''. & The lack of spatial fidelity prevents organic speaker orders like going around in a circle, so meeting leaders often have to decide who speaks when. & \begin{itemize}\vspace{-9pt} \item Display a single ordering of participants that is the same for all users. \item Do not order participants based on whether their cameras are on/off. \item Allow meeting leaders to generate random speaking orders. \end{itemize} \\
\textbf{Raise Hand} & Raise hand is most useful for participants who have trouble interjecting. & Raise hand is only used if the meeting leader explicitly says it should be, but meeting leaders rarely set explicit norms for its use. & \begin{itemize}\vspace{-9pt} \item Provide features for participants to anonymously communicate their preferred raise hand norms. \item Adjust the feature so social expectations are communicated through its technical settings (e.g., by making it toggle-able). \end{itemize} \\
\textbf{Text-based Chat} & Text-based chat lowers barriers to participation, especially for those who might not otherwise participate. & Many participants rely on meeting leaders to incorporate their text-based contributions into the main voice-based conversation. & \begin{itemize}\vspace{-9pt} \item Divert some of text-based chat's cluttering uses to other media (e.g., display status updates on user tiles). \item Allow participants to anonymously nudge meeting leaders towards text-based chat. \end{itemize} \\
\textbf{Meeting Recording} & Meeting recording is a double-edged sword; it negatively impacts some marginalized groups \textit{during} the meeting, but provides crucial access for other marginalized groups \textit{after} the meeting. & Meeting leaders have complete technical control over whether a meeting is recorded, and do not have tools for easily obtaining consent from participants. & \begin{itemize}\vspace{-9pt} \item Provide tools for meeting leaders to obtain recording consent. \item Allow participants to control how they appear on meeting recordings separately from the meetings themselves. \end{itemize}
\end{tabular}
\caption{A summary of our findings in Results and corresponding design recommendations in Discussion. Each recommendation functions through one or both of the mechanisms detailed in Section 5.5---\textbf{transferring control from meeting leaders} and \textbf{helping meeting leaders better exercise their control.}}
\label{tab:summarytable}
\end{table}
}

\usepackage{xcolor}

\AtBeginDocument{%
  \providecommand\BibTeX{{%
    \normalfont B\kern-0.5em{\scshape i\kern-0.25em b}\kern-0.8em\TeX}}}

\begin{document}

\title[``All of the White People Went First'']{``All of the White People Went First'': \\ How Video Conferencing Consolidates Control and Exacerbates Workplace Bias}

\author{Mo Houtti}
\email{houtt001@umn.edu}
\affiliation{
  \institution{University of Minnesota}
  \city{Minneapolis}
  \state{Minnesota}
  \country{USA}
}

\author{Moyan Zhou}
\email{zhou0972@umn.edu}
\affiliation{
  \institution{University of Minnesota}
  \city{Minneapolis}
  \state{Minnesota}
  \country{USA}
}

\author{Loren Terveen}
\email{terveen@umn.edu}
\affiliation{
  \institution{University of Minnesota}
  \city{Minneapolis}
  \state{Minnesota}
  \country{USA}
}

\author{Stevie Chancellor}
\email{steviec@umn.edu}
\affiliation{
  \institution{University of Minnesota}
  \city{Minneapolis}
  \state{Minnesota}
  \country{USA}
}

\renewcommand{\shortauthors}{Mo Houtti, et al.}

\begin{abstract}
Workplace bias creates negative psychological outcomes for employees, permeating the larger organization. Workplace meetings are frequent, making them a key context where bias may occur. Video conferencing (VC) is an increasingly common medium for workplace meetings; we therefore investigated how VC tools contribute to increasing or reducing bias in meetings. Through a semi-structured interview study with 22 professionals, we found that VC features push meeting leaders to exercise control over various meeting parameters, giving leaders an outsized role in affecting bias. We demonstrate this with respect to four core VC features---user tiles, raise hand, text-based chat, and meeting recording---and recommend employing at least one of two mechanisms for mitigating bias in VC meetings---1) transferring control from meeting leaders to technical systems or other attendees and 2) helping meeting leaders better exercise the control they do wield.
\end{abstract}



\maketitle

\section{Introduction}

Research in meeting science has long established the integral role meetings play in organizational functioning, shaping processes, structures, and outcomes~\cite{scott_five_2015}. Meeting quality is an important contributor to the health of many businesses, with the quality of team meetings predicting organizational success up to 2.5 years later~\cite{kauffeld_meetings_2012}. Even putting economic factors aside, meeting quality is immensely relevant to knowledge workers---an estimated 1 billion people globally~\cite{ricard_council_2020}. The typical knowledge worker spends 25-80\% of their working hours in meetings~\cite{romano_meeting_2001}---assuming a 40-hour workweek, this amounts to 10-32 hours \textit{each week}. Employee satisfaction with and participation in meetings is associated with overall job satisfaction and productivity~\cite{zhu_employee_2015, bhatti_impact_2007}, and poor meetings reduce employee well-being, e.g., by increasing emotional exhaustion~\cite{lehmann-willenbrock_our_2016}.

With movements to remote work caused by the COVID-19 pandemic~\cite{levanon_remote_2020}, commercial video conferencing (VC) software providers enjoyed massive increases in the adoption of their products for workplace meetings. Zoom Video Communications---owner and operator of Zoom, the world's most popular video conferencing platform~\cite{brandl_video_2021}---saw its annual revenue more than quadruple from \$623 million in 2020 to \$2.65 billion in 2021~\cite{richter_infographic_2021}. The widespread adoption of computer-mediated communication in an area of such critical importance to individuals and organizations makes professional meetings an area of research with renewed importance in CSCW, and presents us with an opportunity to help answer a question of wide-ranging significance: \textit{how is the adoption of VC likely to change professional meetings?} In exploring this ourselves, we illuminate avenues for improving professional meetings \textit{at scale}; VC is computationally mediated, so we can employ design solutions much more easily, broadly, and cost-effectively than in the equivalent in-person contexts.

To ground this study, we focus our inquiry on a widely occurring problem---that of \textit{bias}. We consider a VC system \textit{biased} if it systematically imposes inequitable costs or barriers to participation on an individual or group---whether on the basis of immutable characteristics such as race, or changing characteristics such as seniority or organizational role.

Our first reason for focusing on bias is pragmatic; compounding the negative outcomes of poor meetings, employees who experience workplace bias are less productive and more likely to quit their jobs earlier~\cite{lekchiri_perceived_2019,moceri_bias_2012}. The growing diversity of the global workforce~\cite{toossi_labor_2015, long_women_2019, kharroubi_global_2021} makes this problem increasingly important.

We also believe addressing bias is important \textit{ethically}. Experiencing bias in the workplace can contribute to depression, anxiety, stress, low self-esteem, and lower overall job satisfaction~\cite{lekchiri_perceived_2019,hughes_african_1997}. Our concern is heightened by the fact that bias disproportionately affects minorities who already face myriad other obstacles at work~\cite{briggs_second_2018,heilman_gender_2007,militello_managing_2015}. We believe VC software designed to decrease bias would be more ethical, and that reducing bias is therefore an inherently worthwhile objective.

The goal of our research was to 1) understand \emph{whether} and \emph{how} VC contributes to bias against groups of users in workplace meetings, and 2) provide specific feature and design recommendations for VC systems to reduce bias in workplace meetings. We conducted semi-structured interviews, asking 22 professionals about bias and barriers to participation they face in workplace meetings related to their use of VC. Through qualitative analysis of the interview transcripts, we provide detailed insight into employees' experiences of bias in meetings, and synthesize these insights to produce concrete design recommendations.

Our results show that VC pushes meeting leaders to take on excess control over core meeting parameters. In some cases, VC does this by directly allocating \textit{technical} control---e.g., a designated meeting host can unilaterally decide whether the meeting is recorded. In other cases, control is allocated \textit{socially}, sometimes because of ambiguities about whether and how various VC features should be used---e.g., whether or not meeting participants should raise their hands to speak. How this control is wielded has major implications for bias in VC meetings. We organize our recommendations for reducing bias around four VC features participants frequently referred to. These features exist in similar forms across most popular VC platforms including Zoom, Cisco Webex, Microsoft Teams, and Google Meet.

We summarize key findings and corresponding design recommendations (bolded) below:

\begin{itemize}

\item \textit{User Tiles.} Inconsistent user tile placement across users' interfaces precludes speaking order from emerging organically, thereby pushing meeting leaders to decide who speaks when. Other aspects of tile placement can lead meeting leaders to call on people in ways that exacerbate already-existing participation inequalities.

\textbf{VC systems should make tile placement across users' interfaces consistent, and/or allow meeting leaders to select random speaking orders.}
\vspace{3mm}

\item \textit{Raise Hand.} Employees rarely use the raise hand feature unless explicitly instructed to do so by a meeting leader. This feature has the potential to lower barriers to participation for people who are uncomfortable interrupting their co-workers.

\textbf{VC systems should enable meeting participants to anonymously communicate their preferred hand-raising norms and/or push meeting leaders to set explicit norms.}
\vspace{3mm}

\item \textit{Text-Based Chat.} Participants who communicate through text-based chat often rely on meeting leaders to acknowledge their questions and contributions to the meeting. Incorporating text-based chat is perceived to reduce bias, and significantly lowers barriers to participation for those who are least likely to speak up.

\textbf{VC systems should divert some of chat's uses (e.g., encouraging messages) to other channels so meeting leaders can more easily parse chat content and incorporate it into meetings.}
\vspace{3mm}

\item \textit{Meeting Recording.} Employees rely on meeting leaders to decide whether a meeting is recorded and how the recording is used. Recording a meeting allows those with with certain cognitive disabilities greater access to meeting content, but may inhibit participation from some minority groups during the meeting. 

\textbf{VC systems should allow meeting participants to hide their appearance in recordings without having to hide their appearance in the meeting} and/or give meeting leaders tools to more easily obtain consent for recording.

\end{itemize}

Not all meetings are led by a single person or group of people, but our participants mostly spoke about meetings where a clear meeting leader could be identified. We expect this is a substantial portion or majority of workplace meetings. We begin our results by outlining the various ways in which our participants identify who the meeting leader is, and set the stage for understanding to whom additional control accrues in meetings. We then outline our findings with regards to the four VC features listed above. Using our participant interviews, we show how each feature as currently implemented in popular commercial VC systems puts control in the hands of the person or people leading the meeting. We then show how this control gives meeting leaders influence over whether and how a meeting is biased. We first, however, outline the literature in which our own work is situated and provide an overview of our methods.

\section{Related Work}

Our work extends research in three main areas. We begin with a description of prior work on VC, focusing specifically on the central challenges CSCW research has tried to address. We then describe the relevant literature on workplace bias, and conclude with an overview of prior work on how meeting leaders shape workplace meeting outcomes.

\subsection{Video Conferencing}

Prior work has documented many advantages of VC use in the workplace and elsewhere, including protecting employees’ health from disease~\cite{okereafor_understanding_2020}, reducing travel expenses and time~\cite{Kydd_managerial_1994}, drawing international participants together from different locations~\cite{thompson_supporting_2009}, and allowing virtual teams to bond socially~\cite{hacker_virtually_2020}.

Our work is situated in the body of HCI and CSCW research that has revealed and improved on VC’s \textit{limitations}, however. 
As Ackerman~\cite{ackerman_intellectual_2000} argued, there is a \textit{``divide between what we know we must support socially and what we can support technically.''} He termed this the \textit{``social-technical gap''}, and argued that \textit{``ameliorating this social–technical gap is the central challenge for CSCW as a field and one of the central problems for human–computer interaction.''} Researchers in CSCW must often provide \textit{approximate} solutions that do not satisfy all the social requirements and social contexts where the technology is used. Olson and Olson~\cite{olson_distance_2000} studied this gap within the context of distance communication technology in particular, and concluded that \textit{``There will likely always be certain kinds of advantages to being together. However, as a wide range of collaborative tools emerges, we will find
useful ways to use them to accomplish our goals.''}

We take on the central intellectual challenge posed by Ackerman, seeking to understand the social-technical gaps that lead VC to exacerbate meeting bias, and to show where the latest approximate solutions lead to bias in workplace meetings. And following Olson and Olson, we continue the line of research that finds useful ways for admittedly imperfect distance communication tools to help us accomplish our goals---in this case, our equity goals. Subsequent work in CSCW has addressed similar gaps in VC's ability to fully support collaborative social interaction.

In a 2016 literature review, Bonaccio et al.~\cite{bonaccio_nonverbal_2016} identified four essential functions of nonverbal communication in the workplace, including promoting social functioning and fostering high-quality relationships. Researchers in HCI and CSCW have found these functions are inhibited in the VC context. Nguyen and Canny~\cite{nguyen_multiview_2007}, for example, compared groups using video conferencing under different conditions and found that traditional VC setups slowed the development of intragroup trust because of limitations on nonverbal gesturing. These adverse effects were minimized, however, in setups that preserved the spatial relationships of participants to each other, allowing participants to gesture towards each other more effectively. Limited nonverbal communication has also been partially credited with the phenomenon of ``Zoom fatigue''---used to describe the physical and emotional exhaustion often experienced by participants in virtual meetings~\cite{bothra_avoid_2020,williams_working_2021}. By building on prior academic theory and research, Bailenson~\cite{bailenson_nonverbal_2021} provided several possible explanations for Zoom fatigue, including that VC requires individuals to put more effort into directing signals for nonverbal cues, resulting in heavier cognitive load.

Our work extends this literature by relating much of it directly to bias in workplace meetings. For example, we find that a shared sense of space is important not only for building trust, but also for collectively making unbiased decisions about speaking order. Spatial fidelity is therefore \textit{also} an important factor in mitigating bias in VC meetings. We also find that VC’s limits on nonverbal communication affect people differently. Many rely on body language to interject in face-to-face workplace meetings, but find that these same tools do not work in the VC-mediated context. When meeting leaders do not explicitly set alternatives for interjecting, marginalized voices are often further marginalized.

\subsection{Workplace Bias}

We take a broad definition of bias in this work, understanding it as encompassing any instance where an inequitable cost or barrier to participation is imposed on a group or individual, intentionally or unintentionally. Prior work has extensively documented contexts in which workplace bias occurs and the characteristics it is frequently based on.

In a literature review of 25 studies, \citet{nowrouzi_weight_2015} observed that bias permeates almost every stage of employment including recruiting, job interviews, hiring, and evaluation. \citet{deprez_accents_2010} found that those making hiring decisions often favored those with accents most similar to their own. \citet{rudolph_meta_2009} found that overweight individuals faced worse outcomes in performance evaluations and hiring decisions. In a literature review of studies on gender discrimination, \citet{badgett_bias_2009} found that 20-57\% of transgender participants reported experiencing workplace discrimination in studies from 1996 to 2006. \citet{dawson_hair_2019} identified Black women whose natural hairstyle choices led them to experience negative workplace consequences.

In CSCW, some attention has been given to bias arising from workplace adoption of VC systems. \citet{das_towards_2021} interviewed 36 neurodivergent individuals about their work-from-home experiences and identified several ways VC software is inaccessible for neurodivergent employees. Some participants, for example, cited difficulties with taking turns speaking given the absence of nonverbal cues. Our own results corroborate this finding, and we suggest potential design improvements.

To our knowledge, there is little other work in CSCW or elsewhere that examines the specific role of VC in workplace bias. This is a particularly important gap because bias that arises primarily from technical systems can be mitigated at scale much more easily than bias that is due to deeply entrenched business practices or human tendencies. We contribute new knowledge here by identifying how features of VC lead to biased meeting outcomes, how these effects differ based on participants' personal characteristics like gender, gender identity, race, neurodivergence, etc., and by providing actionable design recommendations VC software designers can use to mitigate these adverse outcomes.

\subsection{The Role of Meeting Leaders}

Prior work has highlighted the essential role leaders play in meeting dynamics and outcomes. \citet{sauer_ties_2015} performed social network analysis to examine meeting leader behaviors in 46 meeting videos, and found that meeting leaders are key to connecting meeting participants with one another. Several other studies~\cite{schuleigh_effects_2021, schuleigh_enhancing_2019, lehmann_critical_2017, jakobsson_energizing_2021} have identified meeting leaders’ many responsibilities, including planning for the agenda, guiding meeting flow, encouraging participants to exchange ideas, and keeping a healthy meeting environment. In short, meeting leaders play an essential role in meeting success. Given this understanding, Perkins~\cite{perkins_executive_2009} investigated positive meeting leader qualities and how to elicit them via coaching.

Additional attention has been given to leadership in \textit{virtual} meetings in recent years, with much of this work focused on helping leaders run better online meetings. \citet{ferrazzi_run_2015}, for example, created a checklist for best practices meeting leaders can reference before, during, and after meetings: start the meeting with a regular check-in, encourage solving problems collectively, and finish the meeting with a cool ending. Other suggestions include frequently checking attendees’ engagement levels~\cite{kreamer_optimizing_2021}, setting inclusive norms for communication, and utilizing available technology to provide a better meeting experience~\cite{dhawan_videoconferencing_2021}.

Our work extends this literature by showing how many of meeting leaders’ responsibilities arise from VC features pushing them to take control over core meeting parameters, and how this excess responsibility can lead to bias. While we primarily contribute \textit{design recommendations}, a few of our recommendations center around pushing meeting leaders to behave differently (e.g., setting explicit hand-raising norms) and thus, viewed differently, could also be considered recommendations for more effective meeting leadership.

\section{Methods}

Our methods were based on inductive thematic analysis~\cite{braun_thematic_2012}. We now detail our implementation of this approach.

\subsection{Interview Protocol}

Our research team collaboratively developed a semi-structured interview protocol to understand how employees in the workplace experience and perceive bias in VC meetings. The lead author then conducted pilot studies with two graduate students to help refine the interview questions. Data from pilot studies were excluded from our final analysis.

\subsection{Recruitment and Participants}

We submitted our research plan to the University of Minnesota's Institutional Review Board and obtained approval prior to beginning recruitment. We reached out to personal and professional contacts, asking them to circulate our recruitment message in their workplaces. We also used snowball sampling---asking our participants to forward our message to others in their networks. The recruitment message linked to a sign-up form where participants could provide contact information, specify interview availability, list preferred VC platforms, and consent to participation in the study. We used email to schedule VC meetings with those who signed up. All participants listed Zoom as one of their preferred VC platforms, so we used it for all interviews.

After each interview, we used an online survey to collect the participant's demographic information. All questions in the survey were optional, and all except ``job level'' were free-response. Participants’ answers to these survey questions are listed in Table \ref{tab:demographics}. The survey did not ask participants for the industry they worked in, but many participants mentioned their industries during the interviews, which included insurance, educational technology, information technology, and performing arts, in both commercial and higher education settings.

Our initial round of recruitment yielded a participant pool that was diverse in terms of gender, but was largely white and Asian. Given our research topic, we wanted to include a racially diverse set of voices, so we conducted an additional round of recruitment focused on individuals from under-represented minority groups. We reached data saturation shortly before completing our analysis, so we stopped recruiting after the second round. Each participant received a \$20 Amazon gift card as compensation for participating.

\demographics

\subsection{Conducting Interviews}

We began each interview by going over our data privacy practices---e.g., ensuring participants understood they would be anonymized in any related manuscripts/publications, and that we would not disclose their participation to any third parties. We gave participants a chance to ask any remaining questions, then confirmed their consent for us to record the interview.

We explained the general focus of our study. To avoid constraining participant responses to our preconceptions, we did not provide a specific definition of ``bias''. We did refer to ``barriers to participation'' throughout the interviews to ensure participants did consider the type of bias we were focused on. We first asked participants to think of a recent virtual work meeting that seemed relevant to the themes we were concerned about. We then asked specific questions about the meeting setup, followed by more general questions about the interplay of various meeting factors with bias, participation, and engagement---e.g., \emph{``how does camera usage affect your participation in meetings?''} In the spirit of semi-structured interviewing, most of our interview questions flowed from interesting threads brought up by the participants themselves in the course of conversation.

\subsection{Data Collection and Analysis}

The interviews ranged in length from 42:39 to 1:27:13; the average length was 1:04:24. The raw data for our analysis was auto-generated transcripts from the Zoom recordings. The transcripts contained errors, but we could often deduce the original meaning based on context clues and recollection of the interviews. When needed, we also referred back to the original meeting recordings.

We applied open codes on the transcripts using MAXQDA2022---a popular online tool for qualitative data analysis. Two researchers worked together to set standards for open coding on a single interview, then individually generated open codes for the remaining interviews. At the end of this process, we had approximately 1700 open codes, each summarizing a distinct participant statement---e.g., one code read \textit{``P13: Weird to have meeting there in perpetuity esp when misgendered''.}

The same two researchers then performed axial coding together. Using Miro, we iteratively grouped open codes into thematic categories and sub-categories. For example, the aforementioned code for P13 was placed in the sub-category ``permanence'' under the category ``recording''. As overarching patterns became clearer, we altered the categories to better capture the associations between open codes. All results were derived from patterns and themes identified through this process.

\section{Results}

All our participants had extensive experience using VC for workplace meetings. All indicated familiarity with Zoom, though a couple mentioned primarily using Microsoft Teams or Google Meet.

The interviews took place in October through November of 2021, when many of our participants were still working remotely due to COVID-19 pandemic restrictions. Some had little experience with in-person meetings in their current role, either because they had recently started a new job or because they had worked remotely even before the onset of the pandemic. Others had previous in-person meeting experience in the same role, and were therefore able to provide useful comparisons between the two modalities within the same professional context.

Our interviews surfaced many topics and findings from prior academic and popular literature, such as the occurrence of ``Zoom fatigue''~\cite{williams_working_2021,bothra_avoid_2020,hacker_virtually_2020} (P21 described VC as causing \textit{``mental drain''}), work-life conflict caused by having children at home while working remotely~\cite{schieman_work-life_2021} (P7 perceived people who engaged in child-care during meetings \textit{``a little bit less professionally''}), and the VC's ability to increase efficiency through multitasking~\cite{cao_large_2021} (being able to work on other things is \textit{``definitely a net positive''} of VC for P12).

Our focus on bias also revealed how VC pushes or requires meeting leaders to exercise control over core meeting parameters, and how this affects bias in meetings. We summarize these findings and corresponding design recommendations from the Discussion section in Table \ref{tab:summarytable}. As context for this, we first outline who meeting leaders are, according to our participants.

\subsection{Who Leads Meetings?}

Participants indicated that meetings were frequently led by the most senior person or people in attendance. P9, for example, described \textit{``the usual cadence''} as involving \textit{``leadership leading the meeting.''} She recalled one instance of a weekly meeting where \textit{``another person joining the meeting was one level above''} the usual two meeting leaders, \textit{``so the three of them were leading the meeting.''} Leading a meeting often involves setting the agenda, which meeting leaders do with varying levels of structure. In P9's case, the \textit{``two people a level above''} her who were in charge of leading the weekly meeting \textit{``always provide a clear meeting agenda, usually earlier in the day or the day before.''} P10, by contrast, described how her manager runs a weekly team meeting that is \textit{``the most unstructured meeting of my week.''}

\begin{quote}
\textit{Since we have no agenda, and since he doesn't set any boundaries around the meeting or what should happen in the meeting, it's just kind of up to him what he wants to do.}
\end{quote}

There are exceptions to this general rule. In meetings with no clear role hierarchy, for example, meeting leadership could depend on other factors, such as who called the meeting---often the person designated as ``meeting host'' and given special technical privileges:

\begin{quote}
\textit{The host is at the top of the hierarchy, so the host is always allowed to talk, and they can talk to anyone. This is my imagination of how it is; I'm sure it's not written down anywhere.} (P17)
\end{quote}

By extension, having special \textit{technical} privileges to call meetings---and to be meeting host---also confers \textit{social} privileges. P20 called this out explicitly, by observing how her inability to schedule new VC meetings affects how she perceives her role in existing meetings:

\begin{quote}
\textit{I think there are definitely questions of, well, if I don't have access to this platform, I think that impacts how people see their ability to take the lead, or to take up space. So I see that as a dynamic that plays out in the meeting.}
\end{quote}

In some cases, there is no meeting leader at all. P11 mentioned that \textit{``often it's not one specific person leading a meeting,''} especially in cases where the meeting doesn't require a facilitator---e.g., \textit{``it's me and three other people to just brainstorm something.''} (P11) However, participants mostly spoke about meetings where there \textit{was} a clear leader. Our results, therefore, focus on these types of meetings, which we expect encompass a substantial portion---if not a majority---of workplace meetings.

Beyond just setting the agenda and having more space to talk, VC gives meeting leaders new avenues for control over meeting parameters that govern whether and how \textit{others} participate, including in ways that have major implications for bias. We now show how this manifests through particular VC features, beginning with user tiles.

\subsection{User Tiles}

Participants in VC meetings are typically each represented within a tile containing the person's name and, optionally, a video feed from their webcam. Zoom, which the majority of our participants used, has two main views---gallery view, in which participants are given equally sized tiles and organized into rows, and speaker view, in which the current speaker is displayed in a larger tile that takes up most of the screen (Fig. \ref{fig:zoomviews}). Several participants noticed that the order of tiles was not the same on every meeting participant's screen---i.e., the person in the bottom-left tile on one user's screen might be in a different position on another user's screen. While seemingly innocuous, this design decision creates additional reliance on meeting hosts to call on people to speak. We explain how this occurs using an illustrative example from P11's experience. We then outline other aspects of tile positioning that cause hosts to unintentionally exacerbate participation biases.

\zoomviews

\subsubsection{Leader Picks Who Speaks Next}

P11 recounted an instance of bias at a weekly check-in, where her manager called on people individually to share updates. \textit{``He stopped like one or two times, seeming to have to think about it, and then was like `Oh, we didn't hear from whoever yet.'''} P11 noticed that the ad-hoc order in which her manager called on people seemed to be in the \textit{``order of his ranking of importance''}, creating an unintentional but clear bias in favor of those the manager felt closest to:

\begin{quote}
\textit{I wouldn't say it was in order of seniority, but it was almost in order of seniority. But that made it so that all of the white people went first before any of the people of color, which I don't think was intentional, but it was very noticeable to me.}
\end{quote}

The meeting was punctuated by another instance of bias, this time against P11---the only woman present:

\begin{quote}
\textit{And then, when he wrapped up the meeting, I was the only one that he didn't ask to speak. So he forgot about me.}
\end{quote}

P11 speculated about what the same meeting might have looked like in person:

\begin{quote}
\textit{When we were in person, there were similar dynamics at play, but it didn't feel as weird. If we're going around the room doing a check-in, it was always kind of like 'Okay, we'll start here' and there seemed to be a reason, at least, for the order that was picked for people sharing.}
\end{quote}

In other words, being in a physical space together would have allowed meeting participants to use a common practice for assigning speaking order---going around in a circle. But the way in which VC---in this case Zoom---presented each participant with a different physical layout of the meeting precluded them from doing so, causing the meeting leader---P11's manager---to take on additional authority with regards to participation order. P7 noticed the same thing in her meetings, and complained that \textit{``we should all have the same order, because there's always that awkwardness of who's going next.''} She hypothesized that consistent ordering could solve the problem of meeting leaders always \textit{``prioritiz[ing] the people whose videos are on.''} Participants noticed this aspect of tile ordering often meant meeting leaders' decisions about who to call on exacerbated existing participation biases, especially because people who were more likely to speak up were also more likely to have their cameras on. As we will now show, the baseline participation for people with cameras on is greater, and this difference is exacerbated because VC pushes meeting leaders to include those with cameras on more than those with cameras off.

\subsubsection{People With Cameras are Prioritized}

Participants reported higher engagement with camera on, which in turn led them to participate more. P11, for example, noted being \textit{``more engaged if I have my camera on.''} For P22, \textit{``if I have my camera on, I'm usually participating in the meeting more and driving the meeting more.''} Conversely, when P15 has camera off, \textit{``there's nothing coming from me.''}

Meeting leaders and other attendees also reported being more likely to bring participants with cameras on into the conversation. P22, for example, recalled that \textit{``when people have camera on... I'm looking at people and I'm like ‘Oh, this person looks like he or she has something they want to say.'''} Similarly, if P10 \textit{``can tell from their facial expression that they have something to contribute that might be a different opinion, then I would... be like ‘Hey, do you have anything to say? It looks like you haven't said anything. Is everything okay?’ And that just gives them a chance to express their opinion.''} The effect goes both ways, too---having camera off also \textit{actively discourages} bringing a person into the conversation. As P18 described it, there is often \textit{``an implicit understanding that if they're not on video, maybe don't call on that person.''} After all, \textit{``if your camera's off it's probably for good reason---like you're in transit, you're getting your water, you can't necessarily speak up.''} As we show in section \ref{Meeting Recording}, this assumption is often incorrect, as meeting participants frequently turn their cameras off for other reasons---e.g., to avoid being recorded.

Participants noted that visual ordering of users' tiles was based on cameras, further exacerbating the inequalities we outlined above. P7 noticed that the \textit{``type of people who tend to show their camera, they're the same ones that always pop up at the front''} page of her interface. \textit{``It's the same few people who always come up in my default view.''} She criticized Zoom for \textit{``prioritizing the people with the cameras''}---the same people who are already most engaged and likely to speak.

Bringing those with cameras on to the front also involves pushing those with cameras off to the back, leading P3 to observe that \textit{``the people without their cameras on often are treated as being in the back row.''} This is consistent with prior work in proxemics showing that physical positions relative to others in a room affect how likely a person is to be engaged in conversation~\cite{steinzor_spatial_1950}. Thankfully, the raise hand feature provides alternative means by which meeting participants---even those ``in the back row''---could break into the conversation. However, we now outline how even use of this feature largely depends on meeting leaders to set norms for its use.

\subsection{Raise Hand}

The raise hand feature enables participants to indicate they would like to speak. In most current implementations, when a participant raises their virtual hand, they are featured in a more prominent position (e.g., the top-left tile of the first page in gallery view) on other meeting participants' screens, and an icon on their tile indicates a raised hand. Our results indicate that the raise hand feature relies on meeting leaders to set explicit participation norms. Most fail to do so, making the feature chronically under-used in professional VC meetings. This reduces the participation of users who are more hesitant to participate in the first place.

\subsubsection{The Default is to Not Raise Your Hand}

Several participants attributed sparse usage of the raise hand feature to meeting norms. No participants recounted instances of hand-raising being actively discouraged, but several indicated lack of \textit{any} clear guidance around its usage. P14, for example, often wondered, \textit{``Do they want me asking a question in person or should I raise my hand?''} In these cases, participants defaulted to not using the feature. P7, for example, said that raise hand was \textit{``not something we commonly use. It's just not been adopted, so I always feel silly doing it.''} P11 similarly did not raise her hand because \textit{``it's just a norm in this company. I've never seen anyone do it in a non-classroom setting.''} Many decided not to use the feature unless given clear instruction to do so because it did not feel completely appropriate for professional contexts---i.e., it \textit{``seems very elementary school''} (P13) or \textit{``it's almost too awkward.''} (P10) However, in cases where clear guidance to use the raise hand feature \textit{was} provided, participants reported doing so. P10 recalled one such example:

\begin{quote}
\textit{The only time I've had an experience with this was actually last week, where we had a meeting and someone explicitly said, `If you want to say something, please raise your hand.' That's when people started using it.}
\end{quote}

Meeting leaders therefore have often-unexercised (and unrecognized) control over whether the raise hand feature gets used as a tool for interjecting. This has been documented in prior research---especially in the educational context, where hand-raising has been studied the most. \citet{nunneley_leading_2021}, for example, recommended that educators using VC (emphasis ours) \textit{``\textbf{proactively inform their learners how questions should be raised}, such as through chat boxes, `raise hand' features or vocal interjections, and whether they should be spontaneous or asked during specific times.''} As we will now show, this recommendation could apply to the professional meeting context too, as meeting leaders' decisions---or lack thereof---with regards to hand-raising have major implications for whether and how others participate.

\subsubsection{Raise Hand Makes Interjecting Easier}

\textit{``The biggest differentiator''} between the in-person and virtual meeting experiences for P7 was that \textit{``trying to interject myself, I think, is a lot harder in a virtual meeting than in real life.''} Subtle body language cues like \textit{``leaning forward''} (P11) to speak are more difficult to communicate and pick up on. Meetings are often dominated by \textit{``people who aren't as comfortable with silence or who like to speak a lot.''} (P7) They tend to miss the signs that someone is \textit{``trying to break in and say something.''} (P7) This is further exacerbated in cases where participants have their cameras turned off, precluding any reading of facial or body language cues. There is thus a greater burden on people to insert themselves into the conversation, which makes speaking more difficult for people who are hesitant to participate in the first place. \textit{``For remote meetings,''} said P14, \textit{``you have to rely on people bringing themselves forward with more deliberate action to say, `I want to participate.'''}

One such deliberate action could be ``raising'' one's virtual hand, indicating a desire to speak. As P16 put it, it's \textit{``easier in a big group Zoom meeting than a big group in person meeting''} because \textit{``on Zoom, you can just raise your hands.''} Yet as we have shown, participants rarely feel comfortable taking advantage of this feature unless explicitly instructed to do so. This is despite recognizing its benefits for them and others who might not otherwise feel comfortable interjecting. P5, for example:

\begin{quote}
\textit{I don't know if you use [Microsoft] Teams. They have that hand up feature which is useful for like, `Hey I have something to say, but I don't want to interrupt what's going on right now.' I think things like that are really useful.''}
\end{quote}

In keeping with prior work---which has documented how being able to virtually raise a hand makes it easier for students~\cite{noviyanti_students_2022} and conference participants~\cite{smolenskiy_problem_2022} to ask questions---our participants agreed with P5 that the feature is useful for trying to break into an ongoing professional meeting, and lamented that it is not used more. In describing her difficulties with interjecting during meetings, P7 observed that \textit{``some teams, or some people, use the emojis on Zoom to raise their hand''} as a solution. Similarly, P13 is neurodivergent and sometimes has trouble with reading social cues. They do not want to accidentally interrupt someone:

\begin{quote}
\textit{When you're on Zoom, it seems like a lot of social cues just go out the window. I'm not super adept at picking up on social cues in the first place, but I can usually tell when somebody is about to speak---and if I'm about to interrupt them---if I'm in the room with them... I wish people would just raise their hands when they're ready to speak in Zoom meetings... It would help me personally a lot, to feel like I'm not interrupting people, or I'm not speaking over somebody if we just use the tools that are available to us.}
\end{quote}

For P13, this was especially salient in large meetings like their monthly all-hands meetings with over 100 participants:

\begin{quote}
\textit{It seems so silly to me that with that many people, you expect people to just speak up when they're ready to speak... There are people who struggle with being heard in the first place---shy people, neurodivergent people, maybe people of color or trans or LGBT people in general, who are already disadvantaged in these settings where white, cis-gender people speak up more. I think that it makes it even harder for people who are marginalized to participate.}
\end{quote}

Many marginalized meeting participants used text-based chat as an alternative way to have their voices heard, especially when norms about raising hands were ambiguous. P14, for example, described common situations where \textit{``it's unclear when you can speak. You just don't know. Is there a pause? Should I jump in? Do they want me asking a question in person or should I raise my hand? And then you're like, `Uh, I don't know, I'll just type it in.'''} We now cover ways in which the full use of text-based chat depends, once again, on decisions by meeting leaders.

\subsection{Text-Based Chat}

Text-based chat allows users to communicate through text messages---both private messages to a single user and public messages to the entire meeting room---while a meeting is ongoing. Our results show that employees who use text-based chat often rely on meeting leaders to acknowledge their questions and contributions to the meeting, and sometimes feel discouraged from using chat when their contributions are ignored. At the same time, VC makes it difficult for meeting leaders to keep track of and fully incorporate chat, which participants viewed as a secondary medium parallel to the main voice-based conversation. We describe these results, then show how, when successfully used, text-based chat reduces barriers to entry into meeting participation---especially for those who would otherwise be least likely to participate.

\subsubsection{Leader Has to Incorporate Chat Into the Meeting}

Chat was frequently used to ask questions during meetings without interrupting the active speaker. Some participants expressed frustration that presenters often failed to read chat messages. P7, for example, doesn't use chat much because \textit{``sometimes people put a comment in the chat, but by the time someone reads it, you've gone way past the point you were trying to make five minutes ago.''} In addition to questions, some users rely on meeting leaders to acknowledge their contributions and bring them into the main voice-based conversation. P18 described a co-worker for whom this is true:

\begin{quote}
\textit{She uses the chat function a lot. Maybe she's not ready to speak yet, but she'll put her ideas in the chat during this specific meeting I'm referring to, and then maybe the SVP reads it, and if she likes it she'll say, `Oh, I saw that Beth wrote something in the chat.’ And then she's heard that way.}
\end{quote}

P18 added that \textit{``SVPs typically wouldn't use the chat. That medium lends itself to more junior management.''} She hypothesized one reason for higher usage by younger and more junior employees \textit{``could be just comfort with our phones and technology, like growing up with the internet, and these platforms.''} P9's experience offers up an alternative explanation---that keeping track of chat when leading a meeting is just too difficult. \textit{``They're using the chat to ask all these questions and I don't have time to divert from the training and read all of them and answer all of them.''} (P9) As a solution, P9 finds it \textit{``useful to have someone on with me''} who can answer questions in her stead during trainings. This works well for those who ask questions, like P11 who said it \textit{``doesn't particularly matter who answers the question''} as long as someone with the right knowledge does. Some meeting leaders who do not (or cannot) have others serve in this role expressed regret at being unable to incorporate chat messages---P8, for example, lamented that \textit{``they do [use chat], but it's my fault''} for not reading it.

In either case, chat is a medium that is used more by junior meeting participants who typically have less authority in the meeting. But its full use still relies on those leading meetings---usually more senior participants---to acknowledge and incorporate the content therein as necessary. Unfortunately, incorporating chat into the main conversation is difficult, so many meeting leaders fail to do so. When chat use is encouraged and properly incorporated, however, participants reported that it lowered barriers to participation.

\subsubsection{Chat Lowers Barrier to Participation}

As P2 stated, \textit{``being able to just add a little chat comment in there feels really easy. You're not interrupting anyone, you don't have a whole room of people pivoting to look at you in the same way. So I think it feels like there's a lot more psychological safety in being able to add your opinions, your thoughts, or your comments.''} This was most beneficial for those who might otherwise be least likely to participate---e.g., people who are uncomfortable with public speaking. A lot of participation in chat comes from those who \textit{``had there not been a chat, you might never have actually heard them because they might have been too hesitant to actually move into the flow of the conversation.''} (P2) Others echoed this sentiment. P14 hypothesized that big cross-organizational meetings saw more chat use because many were \textit{``nervous to speak in front of all these people.''} P22 similarly expressed appreciation that chat \textit{``allows people to participate, without leaving their comfort zone if they're introverted.''}

All our participants were fluent English-speakers in primarily English-speaking workplaces, but they also noticed how non-native speakers, or native speakers with foreign accents struggled to communicate verbally in professional meetings. P5 recounted how \textit{``we have a couple Indian contractors who speak English fluently but have an accent, and I think sometimes maybe they don't feel as comfortable speaking up in a meeting if they feel like other people can’t understand them well. They have a tendency not to talk in these meetings.''} Chat was presented as a tool that could alleviate the language issues many foreign employees face, in large part because \textit{``at least you don't say the wrong thing if there's a language barrier. You can formulate the question first and make it right before submitting it.''} (P6) While these observations were second-hand---i.e., not from individuals who face language issues themselves---they are consistent with prior work by Satar et al.~\cite{satar_effects_2008}, who found that text-based computer-mediated communication was more effective than voice-based at reducing language anxiety for non-native English speakers.

Not all of chat's use-cases require engagement from the active speakers or presenters. Chat also provides a way for meeting participants to share \textit{``side thoughts''} (P16) without derailing the main voice-based conversation. \textit{``You're not interrupting anyone''} (P2) when you communicate in chat, making it a useful tool for communicating information that is important, but that does not require others to actively engage with it. For example, \textit{``if we’ve been in a meeting for three hours and I really want to use the bathroom''}, P10 uses chat to write \textit{``brb''}. In other cases, parallel chat-based conversations allow substantive discussion about the meeting’s content to occur when they might not otherwise have---for example, if the meeting is \textit{``moving on to the next topic, but they still need to have a conversation about''} (P13) the previous topic.

The ability to communicate without interrupting also brings other benefits, such as frequent expressions of agreement, support, and/or gratitude. P2 remarked that \textit{``unless it's a stump speech at a political rally no one's going to go ‘yeah!’ in the standard business meeting context. But you do get that in chat. ‘Yeah that's a really good idea, I like that point’ or ‘that's interesting.'''} Many participants appreciated that chat enables these positive messages from colleagues. P9, for example, noted that the prevalence of positive messages \textit{``feels like a cool team thing''}. Other participants observed how this was unlikely to work in person or through a voice-based medium. P4, for example, imagined that \textit{``If everyone were to do it, then it would be really loud and sound very horrible''}, whereas text-based chat allows people to \textit{``just type a thank you on the chat''}.

Text-based chat also lets participants skirt the meeting recording feature. We now describe how and why this happens.

\subsection{Meeting Recording} \label{Meeting Recording}

VC software typically allows meeting audio and video to be recorded for later viewing. The ability to start and stop recording a meeting rests entirely with the designated meeting hosts---often the person or people leading the meeting. We find recording meetings to be a double-edged sword that provides crucial access to meeting content for some marginalized groups, while simultaneously reducing the meeting participation of other marginalized groups. Meeting leaders typically make fiat decisions about recording and, we argue, many of them do not sufficiently account for the trade-offs between recording's costs and benefits.

\subsubsection{Meeting Leaders Make Recording Decisions Unilaterally}

Meeting participants are rarely asked for consent to record a meeting. P9, for example, recalled only one such instance:

\begin{quote}
\textit{And then she asked every single person individually in the meeting, `Do you mind if we record this?' And she wanted a verbal `yes' from everyone. And some people are just like `yeah' or they had their camera off and they were on mute and she was like `Robert?', [and I felt] like `this is awkward.' I think that was the only time I experienced that.}
\end{quote}

P9's experience reveals how unusual obtaining consent for recording is, and a potential reason why---VC software provides few features for obtaining consent to record, making it \textit{``awkward''} to do so. Moreover, the power to control meeting recording is typically only given to certain users---often the meeting leaders, and sometimes groups of employees designated by company policy. At P10's company, for example:

\begin{quote}
\textit{One thing to note in our company, we don't have recording rights. I don't have permissions for that. It's the product managers who have those permissions.}
\end{quote}

This concentration of technical power leads other meeting participants to take advantage of other features---e.g., camera and text-based chat---to limit their own appearance on meeting recordings. As we will now show, this alters meeting dynamics and behaviors in ways that are especially salient for certain disadvantaged groups of people---e.g., racial minorities, transgender individuals, and people who have previously had (or currently have) a speech disorder.

\subsubsection{Meeting Recording is a Double-Edged Sword for Bias}

Recording a meeting can have harmful effects that are more pronounced for certain groups of people over others. Many of our participants reported feeling more self-conscious during recorded meetings because they perceived meeting content as permanent. As P21 said, \textit{``you want to make sure, because it's permanent and it's gonna be out there forever, that you come off with the right impression.''} This is consistent with a robust literature comparing ephemeral and persistent online communication---e.g., \citet{bayer_sharing_2016}, who found that Snapchat use comes with reduced concerns about self-presentation because its content is automatically erased after being viewed, and \citet{he_preserving_2020} and \citet{ma_perceived_2021}, who found that users are more comfortable sharing information with others when they perceive it to be ephemeral.

Our participants also described changing their behavior when a meeting was recorded, often by taking advantage of features that \textit{were} in their control. The most common reaction was to turn off their camera which, as we already showed, can lead to reduced engagement and participation. P10, for example, \textit{``wouldn't be as comfortable with keeping my video turned on in a recorded meeting, just because I would be more conscious, and I would have to take care that my expression isn't giving away something.''} P22 also turns off his video when a meeting is being recorded because \textit{``I don't want to be immortalized doing something and having it be a meme. Like picking my nose or something. I mean that's kind of stupid, but I'm being honest here.''} P20 also reported turning off her video when a meeting is recorded, but her concerns centered on her experience as a Black woman in a predominantly white workplace:

\begin{quote}
\textit{Usually if a meeting is recorded, I turn off my camera and will change my name. So as much as I can not show my own image and I can not put my full name, that usually makes me comfortable enough to speak if I want to participate or just be present and take it in. But usually I'm not super excited about video recordings of me, and that comes from a space of constantly having my image used to diversify things. Like they're always going to pick the five most diverse people, however that's determined, and stick their Zoom video squares together and be like `this is the thing we did'.}
\end{quote}

In addition to turning off video, some participants relied more on text-based chat to participate when a meeting was recorded. P10, for example:

\begin{quote}
\textit{When a meeting is being recorded, yes, I do get more conscious. I don't want to unmute myself and say something. I at least gravitate towards the chat where no one's going to hear my voice.}
\end{quote}

P11 recalled behaving similarly:

\begin{quote}
\textit{If it's not recorded, I'll just unmute and answer for him because that's easier. But if it is recorded, I'm way more likely to just put `33' in the chat.}
\end{quote}

This use was informed in part by knowing that chat would be \textit{``hidden when it [the meeting] is recorded.``} (P7) In some cases, however, meeting participants were unable to behave in ways that made them adequately comfortable with meeting recording. The act of recording can, for example, make instances of discrimination more poignant. P13 recalled an uncomfortable situation where they were repeatedly misgendered during a recorded meeting, ensuring the negative experience would be \textit{``there in perpetuity''} rather than being confined to a singular place and time. Moreover, P13 could not take advantage of turning off their camera because doing so made instances of misgendering more likely:

\begin{quote}
\textit{I'm not on hormone replacement therapy at this time, so my voice is coded as female for a lot of people. I think when they hear my voice they automatically react to how it sounds rather than how I look, or what my pronouns are, or how I dress, or whatever. I mean, I always wear men's clothes, you know? I signal to people that I'm trans in other ways than my voice. But I think that, because meetings on Zoom or on the phone are so reliant on how people sound, I think people make a lot of assumptions based on the pitch of my voice, which seems incongruous with my physical appearance or presentation or identity.}
\end{quote}

In other cases, meeting participants cannot take advantage of features like text-based chat because of job expectations---e.g., when giving a presentation. P11 commonly found herself in this situation, and expressed deep discomfort with meeting recordings because of her childhood experience dealing with a speech disorder:

\begin{quote}
\textit{I had a very severe speech impediment growing up, so I don't like public speaking. Can I do it? Sure. Do I want to? Absolutely not, because I hate being on recordings. I don't want there to be videos of me talking.}
\end{quote}

Despite discomfort with meeting recording, our participants generally recognized its value. P6, for example, sees recording as primarily being there to serve \textit{``people who aren't there to listen, that they can watch later.''} This is likely more important for workers who have a greater share of home-life obligations and need greater flexibility---most often women~\cite{coltrane_research_2000}. 

Even when meeting attendance is not an issue, recordings still provide value because they can be consumed in an asynchronous format that allows for speeding up, slowing down, or replaying content. P22 remarked that people often \textit{``talk slower than you can listen to''} so he often finds himself thinking \textit{``I'm so glad this is recorded, because I can go back to it and listen to it on 2x speed.''} Conversely, P20 appreciates that \textit{``for folks with cognitive disabilities that make processing in real time a little bit more challenging, we’re able to record stuff really quick and easy.''} Thus, meeting recordings can increase efficiency while also providing crucial access for individuals who process meetings more slowly, such as those with dyslexia~\cite{stoodley_cerebellum_2011,stoodley_processing_2006} or ADHD~\cite{goth-owens_processing_2010,rommelse_slow_2020}.

However, despite meeting recordings' \textit{potential} utility, many participants expressed skepticism about whether they are \textit{actually} used. P11, for example, hypothesized that \textit{``more often than not, people aren't going to end up watching meeting recordings.''} Other participants expressed the same thing. P12 assumed \textit{``that most people don't even bother to view the recording afterwards,''} but, interestingly, still said \textit{``I prefer not to be recorded.''} P10 echoed this:

\begin{quote}
\textit{I know internally that no one's going to look at it, but it's still something that makes me extra conscious.}
\end{quote}

Discomfort with meeting recordings was not contingent on knowing that the recording would be viewed by others. Participants commonly changed their behaviors in response to meeting recording---often in ways that reduced their meeting participation---but also assumed that the recording would not be viewed. If our participants' speculations are correct, meeting recording's harms \textit{during} meetings are not likely to be balanced out by its benefits for others \textit{after} meetings---i.e., recording's \textit{potential} benefits do not \textit{actually} accrue due to lack of use. This leads us to believe that many meeting leaders do not adequately engage with meeting recording's trade-offs.

\summarytable

\section{Discussion}

We begin our discussion by focusing on the four features of VC, synthesizing prior work alongside our results to describe how each feature contributes to bias. Then, we outline strategies VC system designers and future researchers can implement or explore to mitigate bias in VC systems for each feature, summarized in Table \ref{tab:summarytable}.

\subsection{User Tiles}

\subsubsection{Lack of Spatial Fidelity Makes Leaders Decide Who Speaks}

CSCW researchers have long understood that the lack of shared physical space hinders spatial referencing in VC meetings---e.g., directing your gaze or pointing at someone~\cite{benford_shared_1996}. Our results build on these findings by showing that spatial fidelity also influences prioritization in meetings. As P7 and P11 indicated, face-to-face meeting participants use a shared sense of space to make collective decisions about speaking order and priority. The inconsistent ordering of user tiles precludes this from happening, requiring meeting leaders to decide who speaks when. While some VC systems (e.g., Zoom) allow for a common tile ordering to be set by the host, none of our participants reported using this feature and, in fact, none of this paper's authors even knew the feature existed during interviews. Participants must again rely on meeting leaders to take proactive steps whose benefits are unclear, unless perhaps they have read (or written) this paper.

\subsubsection{Tile Positioning Makes Likely Speakers More Visible}

When given the responsibility to choose who speaks, participants report that leaders decide in systematically biased ways. Conceptualizing tile placement as a virtual form of \textit{seating arrangement} allows us to draw from prior work in proxemics to understand how VC causes this. Steinzor~\cite{steinzor_spatial_1950} found that people in small conversational circles tend to speak to those immediately in front of them more than those by their sides. VC \textit{systematically} positions those with cameras on---those who are already most likely to speak up---directly in front of meeting leaders, thereby further increasing the likelihood they will be engaged in conversation. VC can \textit{further} exacerbate these biases because it is not constrained by physical limitations that apply in the face-to-face context, such as when 10 people occupy the first page on every meeting participant's interface because they have their cameras on. An analogous situation in person would be if everyone in the room were facing these 10 people, \textit{including the 10 people themselves}. While impossible in a \textit{physical} space, this is exactly the situation meeting participants are often confronted with in the VC-mediated \textit{virtual} space.

\subsubsection{How to Mitigate Bias}

There are simple solutions to problems with visibility and tile order. First, VC systems could stop visually prioritizing the tiles of users with their cameras on, opting for a random arrangement instead. Future work should also explore other strategies for arranging user tiles; ordering participants by ascending order of speaking time could help equalize participation by making those who have spoken less more visible (and vice versa). VC systems could also make user tile ordering consistent between participants' screens, like Zoom has in the aforementioned buried feature. Alternatively, VC systems could add tools that let meeting leaders generate random speaking orders, thereby outsourcing decisions about who speaks next to transparent technical systems.

\subsection{Raise Hand}

\subsubsection{Raise Hand Makes Meeting Participation Easier}

Prior work has explored VC's raise hand feature in educational contexts, where hand-raising is more common. \citeauthor{noviyanti_students_2022}~\cite{noviyanti_students_2022} and \citeauthor{smolenskiy_problem_2022}~\cite{smolenskiy_problem_2022} have documented how the raise hand features makes it easier for students and conference participants, respectively, to ask questions. Our findings confirm and extend this literature, suggesting it is generalizable to professional meetings. P7 and P14 said that interjecting in VC meetings is more difficult because participants cannot effectively make use of body language cues. This is especially true for participants who have their cameras off. P8 also observed that raise hand is especially useful for people whose voices are marginalized, making the feature useful for reducing bias in meetings.

~\subsubsection{VC Enables Ambiguous Hand Raising Norms}

Once again, however, the raise hand feature's ability to mitigate bias requires meeting leaders to encourage its use. Unless meeting leaders \textit{explicitly} encourage meeting participants to use the raise hand feature, our participants said it was typically not used. At first glance, this seems to be a social problem with a ``straightforward'' social solution---tell meeting leaders to set explicit participation norms about raise hand. Indeed, this aligns with the existing literature that gives advice on how to better manage meetings (e.g.,~\cite{ferrazzi_run_2015, kreamer_optimizing_2021,dhawan_videoconferencing_2021}). However, the UI implementation of this feature in most VC systems creates normative ambiguity around hand-raising. For example, the raise hand feature is often available by default for meeting participants to use---even in small meetings where participants might agree it \textit{should not} or will not be used. The availability of the feature, therefore, does not signal anything about its expected usage in meeting participation.

\subsubsection{How to Mitigate Bias}

Bias could be mitigated by making hand-raising norms clearer. This could be done with UIs where \textit{technical} settings communicate \textit{social} expectations, in what Erickson and Kellogg would classically describe as ``social translucence''~\cite{erickson_social_2000}. VC systems could implement a feature where raise hand was off by default, and meeting participants could anonymously request that the meeting leader toggle the feature. The \textit{technical} ability for meeting participants to raise their hands might become interpreted as a \textit{social} invitation to raise them, while also giving meeting participants a low-cost way to ask meeting leaders to set participation norms.

\subsection{Text-based Chat}

\subsubsection{Low Cost of Chat Can Reduce Meeting Bias}

Many of text-based chat's benefits in our results were because it enabled participants to communicate without interrupting. This corroborates prior work by ~\citet{scholl_comparison_2006} who found that users in a multi-media communication environment (video, audio, and chat) often preferred communicating through chat because it felt less intrusive. ~\citet{scholl_comparison_2006} suggested this quality could make chat useful for supporting informal interactions at work because of its low personal cost for the receiver---i.e., it does not involve interrupting and can be ignored.

Our findings reveal that this quality of also makes text-based chat useful for reducing bias in meetings. Many employees seem to prefer low-cost interactions \textit{even when} communication is formal and work-related, and many of them \textit{do not contribute at all} unless given a low-cost alternative to the main voice-based channel. In VC meetings, however, employees do \textit{not} appreciate that their chat messages can be---and often are---ignored, so meeting leaders are saddled with greater responsibility to engage with chat.

\subsubsection{VC Makes Chat Difficult to Incorporate}

When meeting leaders fail to adequately incorporate chat, some participants---like P7---stop using it altogether, and its bias-reducing potential remains unrealized. Chat's implementation in contemporary VC systems contributes to this. On top of all their regular responsibilities in running a meeting, meeting leaders are expected to keep track of two separate communication channels simultaneously. As P22 mentioned, chat frequently becomes chaotic because it is used for multiple purposes---questions, comments, side-conversations, words of encouragement, etc. Some---like P9---get around this by enlisting help from a meeting co-facilitator whose sole responsibility is to keep track of chat. This is a costly solution, however, and the need for it could be mitigated by \textit{technical} solutions, which scale much more efficiently.

\subsubsection{How to Mitigate Bias}

Making it easier for meeting leaders to incorporate chat in their meetings can reduce bias. VC systems could do this by enabling meeting participants to ``push'' messages to the main video feed, or anonymously nudge the meeting leader to look at chat. This would let participants make their low-cost communications more visible.

VC systems could also reduce clutter by separating out different types of chat messages. For example, words of praise could be automatically detected and ephemerally presented elsewhere---such as a corner of the main video feed---to prevent them from drowning out questions and other substantive contributions to the meeting. Adding user status icons or custom status text---e.g., ``stepping away for a quick bathroom break'' on a participant's user tile---could also divert some of chat's usage to other media, leaving more room for substantive contributions to be noticed.

\subsection{Meeting Recording}

\subsubsection{Meeting Recording Affects Groups Differently}

Our findings on meeting recording extend a robust literature showing that users are more comfortable communicating through ephemeral media~\cite{bayer_sharing_2016, he_preserving_2020, ma_perceived_2021}. We similarly find that VC meeting attendees are more comfortable participating when their communication is ephemeral, and less so when it is persistent in a meeting recording.

The act of recording a meeting can therefore increase participation bias, because members of certain marginalized groups---e.g., P13, who was repeatedly misgendered in a recorded meeting---are affected more adversely by permanence. Conversely, meeting recording provides a means by which other marginalized groups---e.g., people with certain cognitive disabilities, as P20 mentioned---can access otherwise inaccessible meeting content. Recording is a double-edged sword and, once again, the onus for reaping its benefits and minimizing its costs lay almost entirely with meeting leaders.

\subsubsection{Recording is All-or-Nothing With Few Tools for Consent}

To understand whether recording is appropriate, meeting leaders must know whether the recording is likely to be viewed later, and how individual meeting participants are likely to feel about being recorded. Meeting leaders are not likely to have this information, yet VC systems commonly give them unilateral and \textit{inalienable} control over recording.

As P9 recounted, getting consent to record is difficult and awkward, so it rarely happens. Some VC systems like Zoom can be set to notify participants when recording begins, allowing anyone not consenting to leave the meeting. In practice, however, leaving a recorded meeting is rarely a viable option. And as we have already shown, turning off video is a sub-optimal solution that reduces the person's visibility in the meeting---not \textit{just} in the recording.

Unfortunately, recording in many VC systems is all-or-nothing; meeting hosts must record the entirety of a meeting's audio \textit{and} video stream or record nothing at all. In some systems, the option to record audio only exists, but is often buried in a separate interface. Participants must therefore once again rely on meeting leaders to make use of the available features.

\subsubsection{How to Mitigate Bias}

VC systems could reduce bias from recording by allowing meeting leaders to distribute control to other meeting participants. A feature that enables meeting leaders to quickly and efficiently obtain individual consent from users could encourage communication within organizations about recording's costs and benefits. Giving meeting participants more control over how they appear in recordings could also help mitigate participation bias. Several participants---e.g., P20, P21, and P22---either turned off their cameras or were more conscious of their physical appearances when a meeting was recorded. Enabling them to keep their cameras on during the meeting, but prevent themselves from showing up in the meeting \textit{recording} could have perhaps assuaged their concerns without requiring any changes to their in-meeting behaviors.

\subsection{How to Mitigate Bias: Two Primary Mechanisms}

We find that VC systems give leaders excess control over key meeting parameters that affect bias. Meeting leaders are unable to delegate their control, ill-equipped to exercise it properly, or both, so our recommendations for how to mitigate bias function through at least one of two mechanisms---\textbf{transferring control from meeting leaders} to technical systems or other attendees and \textbf{helping meeting leaders better exercise the control they do wield}. These are central to each of our previous recommendations, so we find this is a useful framework for bias mitigation in VC systems regardless of the specific feature being considered.

For example, our recommendation to make tile ordering consistent between users aims to \textbf{transfer control from meeting leaders} to the group as a whole with regards to speaking order. Our recommendation that VC systems enable participants to alter how they appear on meeting recordings follows from the same thinking about transferring control, despite its pertinence to a different feature entirely. 

The same is true of the latter mechanism. VC systems can reduce bias by enabling meeting participants to anonymously provide information about their preferred hand-raising norms, or by making text-based chat less chaotic. In both cases, the aim is to \textbf{help meeting leaders exercise their control}---over hand-raising norms and text-based chat, respectively---in ways that enable broader participation.

Selecting a mechanism (or combination of mechanisms) for a specific context must consider the cultural and contextual values of the organization. If reducing power imbalances is a primary goal, then transferring control from meeting leaders might be preferred. On the other hand, if the organizational culture is more hierarchical and deferential to authority, this may be difficult, and the emphasis may lean towards helping leaders exercise their control more effectively.

VC systems' failure to leverage one or the other mechanism can persistently reinforce the marginalization of already-marginalized voices. We now outline how the marginalizing effects of one feature could compound with or mitigate those from other features.

\subsection{Compounding or Mitigating Effects}

Let us imagine a hypothetical employee, P0, and consider the following example in which all her routes to participation are suppressed:

\begin{itemize}
\item \textbf{Recording.} P0 decides to turn off her camera to avoid showing up in the meeting recording. This causes her to become less engaged.

\item \textbf{User Tiles.} P0 would like to participate, but turning her camera off has made her less visible to the meeting leader, so she is not brought into the conversation.

\item \textbf{Raise Hand.} P0 considers raising her hand to make a comment, but the meeting leader has not set clear norms around raising hands. She decides not to because it feels silly.

\item \textbf{Text-Based Chat.} P0 decides to put her comment in chat, but the meeting leader does not see it until 10 minutes later, by which time she has already tuned out of the meeting completely. She decides to stop using chat because it doesn't seem like an effective way to contribute.
\end{itemize}

At each step, our imagined participant's marginalization results from a meeting leader having too much control and/or not having the right tools to exercise their control appropriately. While hypothetical, this example illustrates how each of these features could plausibly work together to further marginalize participants who are already marginalized by one of them. Conversely, features such as text-based chat and raise hand, when used properly, can also help mitigate biases that arise from meeting recording and issues with tile placement. For example, had the meeting leader acknowledged P0's chat message, P0 would have been able to contribute to the meeting in some form despite the barriers imposed by meeting recording, tile placement, and ambiguous norms around raising hands. Notably, these barriers arise even in cases where the meeting leader fully intends to create an equitable meeting environment---i.e., it is in large part due to how VC makes meeting leaders assume control over meeting parameters they would otherwise not have to in person, but does not communicate any of these new expectations.

\subsection{Limitations}

We now outline some of our work's limitations. First, we focused primarily on VC meetings in English-speaking (almost all US-based) workplaces, and only interviewed employees who were fluent in English. Many of our findings may not apply to other contexts or cultures, and may not provide a complete picture with regards to the experience of non-native English speakers in English-speaking workplaces. For example, our findings with regards to text-based chat's perceived benefits for non-native speakers, while consistent with prior work, cannot serve as concrete evidence of chat's benefits for this group. Similarly, our use of snowball sampling limits the types of jobs and workers represented. For example, while our participants occupied a variety of industries, many of them worked at tech companies or in technical departments at non-tech companies.

Our inquiry also had broad aims. We did not focus on a single type of meeting or---perhaps most importantly---a single population of marginalized individuals. More specific work that builds on our findings---for example, work that investigates the experiences of Black employees with meeting recording, or how transgender individuals use VC features to communicate their identities---could contribute richness and depth we do not provide here.

\section{Conclusion}

This work contributes empirical findings that reveal how VC systems create bias in workplace meetings. We show that bias in VC often results from excess control being pushed onto meeting leaders. Our design recommendations emphasize two important mechanisms for reducing bias---transferring control from meeting leaders and helping meeting leaders better exercise their control. For both pragmatic and ethical reasons, we hope the knowledge provided here will help move VC systems towards designs that produce more equitable outcomes in workplace meetings.

\begin{acks}
The work reported in this paper was supported by Cisco Research. We would also like to acknowledge and thank our participants for helping to make this work possible.
\end{acks}

\bibliographystyle{ACM-Reference-Format}
\bibliography{main}


\begin{thebibliography}{64}


\ifx \showCODEN    \undefined \def \showCODEN     #1{\unskip}     \fi
\ifx \showDOI      \undefined \def \showDOI       #1{#1}\fi
\ifx \showISBNx    \undefined \def \showISBNx     #1{\unskip}     \fi
\ifx \showISBNxiii \undefined \def \showISBNxiii  #1{\unskip}     \fi
\ifx \showISSN     \undefined \def \showISSN      #1{\unskip}     \fi
\ifx \showLCCN     \undefined \def \showLCCN      #1{\unskip}     \fi
\ifx \shownote     \undefined \def \shownote      #1{#1}          \fi
\ifx \showarticletitle \undefined \def \showarticletitle #1{#1}   \fi
\ifx \showURL      \undefined \def \showURL       {\relax}        \fi
\providecommand\bibfield[2]{#2}
\providecommand\bibinfo[2]{#2}
\providecommand\natexlab[1]{#1}
\providecommand\showeprint[2][]{arXiv:#2}

\bibitem[Ackerman(2000)]%
        {ackerman_intellectual_2000}
\bibfield{author}{\bibinfo{person}{Mark~S. Ackerman}.}
  \bibinfo{year}{2000}\natexlab{}.
\newblock \showarticletitle{The intellectual challenge of {CSCW}: the gap
  between social requirements and technical feasibility}.
\newblock \bibinfo{journal}{\emph{Human-Computer Interaction}}
  \bibinfo{volume}{15}, \bibinfo{number}{2} (\bibinfo{date}{Sept.}
  \bibinfo{year}{2000}), \bibinfo{pages}{179--203}.
\newblock
\showISSN{0737-0024}
\urldef\tempurl%
\url{https://doi.org/10.1207/S15327051HCI1523_5}
\showDOI{\tempurl}


\bibitem[Badgett et~al\mbox{.}(2009)]%
        {badgett_bias_2009}
\bibfield{author}{\bibinfo{person}{M.V.~Lee Badgett}, \bibinfo{person}{Brad
  Sears}, \bibinfo{person}{Holning Lau}, {and} \bibinfo{person}{Deborah Ho}.}
  \bibinfo{year}{2009}\natexlab{}.
\newblock \showarticletitle{Bias in the workplace: Consistent evidence of
  sexual orientation and gender identity discrimination 1998-2008}.
\newblock \bibinfo{journal}{\emph{Chicago-Kent Law Review}}
  \bibinfo{volume}{84} (\bibinfo{year}{2009}), \bibinfo{pages}{559}.
\newblock


\bibitem[Bailenson(2021)]%
        {bailenson_nonverbal_2021}
\bibfield{author}{\bibinfo{person}{Jeremy~N Bailenson}.}
  \bibinfo{year}{2021}\natexlab{}.
\newblock \showarticletitle{Nonverbal overload: A theoretical argument for the
  causes of Zoom fatigue}.
\newblock  (\bibinfo{year}{2021}).
\newblock


\bibitem[Bayer et~al\mbox{.}(2016)]%
        {bayer_sharing_2016}
\bibfield{author}{\bibinfo{person}{Joseph~B. Bayer}, \bibinfo{person}{Nicole~B.
  Ellison}, \bibinfo{person}{Sarita~Y. Schoenebeck}, {and}
  \bibinfo{person}{Emily~B. Falk}.} \bibinfo{year}{2016}\natexlab{}.
\newblock \showarticletitle{Sharing the small moments: ephemeral social
  interaction on {Snapchat}}.
\newblock \bibinfo{journal}{\emph{Information, Communication \& Society}}
  \bibinfo{volume}{19}, \bibinfo{number}{7} (\bibinfo{date}{July}
  \bibinfo{year}{2016}), \bibinfo{pages}{956--977}.
\newblock
\showISSN{1369-118X}
\urldef\tempurl%
\url{https://doi.org/10.1080/1369118X.2015.1084349}
\showDOI{\tempurl}
\newblock
\shownote{Publisher: Routledge \_eprint:
  https://doi.org/10.1080/1369118X.2015.1084349}.


\bibitem[Benford et~al\mbox{.}(1996)]%
        {benford_shared_1996}
\bibfield{author}{\bibinfo{person}{Steve Benford}, \bibinfo{person}{Chris
  Brown}, \bibinfo{person}{Gail Reynard}, {and} \bibinfo{person}{Chris
  Greenhalgh}.} \bibinfo{year}{1996}\natexlab{}.
\newblock \showarticletitle{Shared spaces: transportation, artificiality, and
  spatiality}. In \bibinfo{booktitle}{\emph{Proceedings of the 1996 {ACM}
  conference on {Computer} supported cooperative work}}
  \emph{(\bibinfo{series}{{CSCW} '96})}. \bibinfo{publisher}{Association for
  Computing Machinery}, \bibinfo{address}{New York, NY, USA},
  \bibinfo{pages}{77--86}.
\newblock
\showISBNx{978-0-89791-765-0}
\urldef\tempurl%
\url{https://doi.org/10.1145/240080.240196}
\showDOI{\tempurl}


\bibitem[Bhatti and Qureshi(2007)]%
        {bhatti_impact_2007}
\bibfield{author}{\bibinfo{person}{Komal~Khalid Bhatti} {and}
  \bibinfo{person}{Tahir~Masood Qureshi}.} \bibinfo{year}{2007}\natexlab{}.
\newblock \showarticletitle{Impact {Of} {Employee} {Participation} {On} {Job}
  {Satisfaction}, {Employee} {Commitment} {And} {Employee} {Productivity}}.
\newblock  (\bibinfo{date}{June} \bibinfo{year}{2007}), \bibinfo{pages}{16}.
\newblock


\bibitem[Bonaccio et~al\mbox{.}(2016)]%
        {bonaccio_nonverbal_2016}
\bibfield{author}{\bibinfo{person}{Silvia Bonaccio}, \bibinfo{person}{Jane
  O’Reilly}, \bibinfo{person}{Sharon~L. O’Sullivan}, {and}
  \bibinfo{person}{François Chiocchio}.} \bibinfo{year}{2016}\natexlab{}.
\newblock \showarticletitle{Nonverbal {Behavior} and {Communication} in the
  {Workplace}: {A} {Review} and an {Agenda} for {Research}}.
\newblock \bibinfo{journal}{\emph{Journal of Management}} \bibinfo{volume}{42},
  \bibinfo{number}{5} (\bibinfo{date}{July} \bibinfo{year}{2016}),
  \bibinfo{pages}{1044--1074}.
\newblock
\showISSN{0149-2063}
\urldef\tempurl%
\url{https://doi.org/10.1177/0149206315621146}
\showDOI{\tempurl}
\newblock
\shownote{Publisher: SAGE Publications Inc}.


\bibitem[Bothra(2020)]%
        {bothra_avoid_2020}
\bibfield{author}{\bibinfo{person}{Sweta Bothra}.}
  \bibinfo{year}{2020}\natexlab{}.
\newblock \bibinfo{title}{How to avoid Zoom fatigue while working from home}.
\newblock
\newblock


\bibitem[Brandl(2021)]%
        {brandl_video_2021}
\bibfield{author}{\bibinfo{person}{Robert Brandl}.}
  \bibinfo{year}{2021}\natexlab{}.
\newblock \bibinfo{title}{The {Video} {Call} {Platforms} that {Dominate} the
  {World}}.
\newblock
\newblock
\urldef\tempurl%
\url{https://www.emailtooltester.com/en/blog/video-conferencing-market-share/}
\showURL{%
\tempurl}


\bibitem[Braun and Clarke(2012)]%
        {braun_thematic_2012}
\bibfield{author}{\bibinfo{person}{Virginia Braun} {and}
  \bibinfo{person}{Victoria Clarke}.} \bibinfo{year}{2012}\natexlab{}.
\newblock \showarticletitle{Thematic analysis}.
\newblock In \bibinfo{booktitle}{\emph{{APA} handbook of research methods in
  psychology, {Vol} 2: {Research} designs: {Quantitative}, qualitative,
  neuropsychological, and biological}}. \bibinfo{publisher}{American
  Psychological Association}, \bibinfo{address}{Washington, DC, US},
  \bibinfo{pages}{57--71}.
\newblock
\showISBNx{978-1-4338-1005-3}
\urldef\tempurl%
\url{https://doi.org/10.1037/13620-004}
\showDOI{\tempurl}


\bibitem[Briggs(2018)]%
        {briggs_second_2018}
\bibfield{author}{\bibinfo{person}{Anthony~Q. Briggs}.}
  \bibinfo{year}{2018}\natexlab{}.
\newblock \showarticletitle{Second generation {Caribbean} black male youths
  discuss obstacles to educational and employment opportunities: a critical
  race counter-narrative analysis}.
\newblock \bibinfo{journal}{\emph{Journal of Youth Studies}}
  \bibinfo{volume}{21}, \bibinfo{number}{4} (\bibinfo{date}{April}
  \bibinfo{year}{2018}), \bibinfo{pages}{533--549}.
\newblock
\showISSN{1367-6261}
\urldef\tempurl%
\url{https://doi.org/10.1080/13676261.2017.1394997}
\showDOI{\tempurl}
\newblock
\shownote{Publisher: Routledge \_eprint:
  https://doi.org/10.1080/13676261.2017.1394997}.


\bibitem[Cao et~al\mbox{.}(2021)]%
        {cao_large_2021}
\bibfield{author}{\bibinfo{person}{Hancheng Cao}, \bibinfo{person}{Chia-Jung
  Lee}, \bibinfo{person}{Shamsi Iqbal}, \bibinfo{person}{Mary Czerwinski},
  \bibinfo{person}{Priscilla N~Y Wong}, \bibinfo{person}{Sean Rintel},
  \bibinfo{person}{Brent Hecht}, \bibinfo{person}{Jaime Teevan}, {and}
  \bibinfo{person}{Longqi Yang}.} \bibinfo{year}{2021}\natexlab{}.
\newblock \showarticletitle{Large {Scale} {Analysis} of {Multitasking}
  {Behavior} {During} {Remote} {Meetings}}. In
  \bibinfo{booktitle}{\emph{Proceedings of the 2021 {CHI} {Conference} on
  {Human} {Factors} in {Computing} {Systems}}} \emph{(\bibinfo{series}{{CHI}
  '21})}. \bibinfo{publisher}{Association for Computing Machinery},
  \bibinfo{address}{New York, NY, USA}, \bibinfo{pages}{1--13}.
\newblock
\showISBNx{978-1-4503-8096-6}
\urldef\tempurl%
\url{https://doi.org/10.1145/3411764.3445243}
\showDOI{\tempurl}


\bibitem[Coltrane(2000)]%
        {coltrane_research_2000}
\bibfield{author}{\bibinfo{person}{Scott Coltrane}.}
  \bibinfo{year}{2000}\natexlab{}.
\newblock \showarticletitle{Research on {Household} {Labor}: {Modeling} and
  {Measuring} the {Social} {Embeddedness} of {Routine} {Family} {Work}}.
\newblock \bibinfo{journal}{\emph{Journal of Marriage and Family}}
  \bibinfo{volume}{62}, \bibinfo{number}{4} (\bibinfo{year}{2000}),
  \bibinfo{pages}{1208--1233}.
\newblock
\showISSN{1741-3737}
\urldef\tempurl%
\url{https://doi.org/10.1111/j.1741-3737.2000.01208.x}
\showDOI{\tempurl}
\newblock
\shownote{\_eprint:
  https://onlinelibrary.wiley.com/doi/pdf/10.1111/j.1741-3737.2000.01208.x}.


\bibitem[Das et~al\mbox{.}(2021)]%
        {das_towards_2021}
\bibfield{author}{\bibinfo{person}{Maitraye Das}, \bibinfo{person}{John Tang},
  \bibinfo{person}{Kathryn~E Ringland}, {and} \bibinfo{person}{Anne~Marie
  Piper}.} \bibinfo{year}{2021}\natexlab{}.
\newblock \showarticletitle{Towards accessible remote work: Understanding
  work-from-home practices of neurodivergent professionals}.
\newblock \bibinfo{journal}{\emph{Proceedings of the ACM on Human-Computer
  Interaction}} \bibinfo{volume}{5}, \bibinfo{number}{CSCW1}
  (\bibinfo{year}{2021}), \bibinfo{pages}{1--30}.
\newblock


\bibitem[Dawson et~al\mbox{.}(2019)]%
        {dawson_hair_2019}
\bibfield{author}{\bibinfo{person}{Gail~A Dawson}, \bibinfo{person}{Katherine~A
  Karl}, {and} \bibinfo{person}{Joy~V Peluchette}.}
  \bibinfo{year}{2019}\natexlab{}.
\newblock \showarticletitle{Hair matters: Toward understanding natural black
  hair bias in the workplace}.
\newblock \bibinfo{journal}{\emph{Journal of Leadership \& Organizational
  Studies}} \bibinfo{volume}{26}, \bibinfo{number}{3} (\bibinfo{year}{2019}),
  \bibinfo{pages}{389--401}.
\newblock


\bibitem[Deprez-Sims and Morris(2010)]%
        {deprez_accents_2010}
\bibfield{author}{\bibinfo{person}{Anne-Sophie Deprez-Sims} {and}
  \bibinfo{person}{Scott~B Morris}.} \bibinfo{year}{2010}\natexlab{}.
\newblock \showarticletitle{Accents in the workplace: Their effects during a
  job interview}.
\newblock \bibinfo{journal}{\emph{International Journal of Psychology}}
  \bibinfo{volume}{45}, \bibinfo{number}{6} (\bibinfo{year}{2010}),
  \bibinfo{pages}{417--426}.
\newblock


\bibitem[Dhawan et~al\mbox{.}(2021)]%
        {dhawan_videoconferencing_2021}
\bibfield{author}{\bibinfo{person}{Natasha Dhawan}, \bibinfo{person}{Molly
  Carnes}, \bibinfo{person}{Angela Byars-Winston}, {and}
  \bibinfo{person}{Narjust Duma}.} \bibinfo{year}{2021}\natexlab{}.
\newblock \showarticletitle{Videoconferencing Etiquette: Promoting Gender
  Equity During Virtual Meetings}.
\newblock \bibinfo{journal}{\emph{Journal of Women's Health}}
  \bibinfo{volume}{30}, \bibinfo{number}{4} (\bibinfo{year}{2021}),
  \bibinfo{pages}{460--465}.
\newblock


\bibitem[Erickson and Kellogg(2000)]%
        {erickson_social_2000}
\bibfield{author}{\bibinfo{person}{Thomas Erickson} {and}
  \bibinfo{person}{Wendy~A. Kellogg}.} \bibinfo{year}{2000}\natexlab{}.
\newblock \showarticletitle{Social translucence: an approach to designing
  systems that support social processes}.
\newblock \bibinfo{journal}{\emph{ACM Transactions on Computer-Human
  Interaction}} \bibinfo{volume}{7}, \bibinfo{number}{1} (\bibinfo{date}{March}
  \bibinfo{year}{2000}), \bibinfo{pages}{59--83}.
\newblock
\showISSN{1073-0516}
\urldef\tempurl%
\url{https://doi.org/10.1145/344949.345004}
\showDOI{\tempurl}


\bibitem[Ferrazzi(2015)]%
        {ferrazzi_run_2015}
\bibfield{author}{\bibinfo{person}{Keith Ferrazzi}.}
  \bibinfo{year}{2015}\natexlab{}.
\newblock \showarticletitle{How to run a great virtual meeting}.
\newblock \bibinfo{journal}{\emph{Harvard Business Review}}
  (\bibinfo{year}{2015}), \bibinfo{pages}{2--5}.
\newblock


\bibitem[Goth-Owens et~al\mbox{.}(2010)]%
        {goth-owens_processing_2010}
\bibfield{author}{\bibinfo{person}{Timothy~L. Goth-Owens},
  \bibinfo{person}{Cecilia Martinez-Torteya}, \bibinfo{person}{Michelle~M.
  Martel}, {and} \bibinfo{person}{Joel~T. Nigg}.}
  \bibinfo{year}{2010}\natexlab{}.
\newblock \showarticletitle{Processing {Speed} {Weakness} in {Children} and
  {Adolescents} with {Non}-{Hyperactive} but {Inattentive} {ADHD} ({ADD})}.
\newblock \bibinfo{journal}{\emph{Child Neuropsychology}} \bibinfo{volume}{16},
  \bibinfo{number}{6} (\bibinfo{date}{Nov.} \bibinfo{year}{2010}),
  \bibinfo{pages}{577--591}.
\newblock
\showISSN{0929-7049}
\urldef\tempurl%
\url{https://doi.org/10.1080/09297049.2010.485126}
\showDOI{\tempurl}
\newblock
\shownote{Publisher: Routledge \_eprint:
  https://doi.org/10.1080/09297049.2010.485126}.


\bibitem[Hacker et~al\mbox{.}(2020)]%
        {hacker_virtually_2020}
\bibfield{author}{\bibinfo{person}{Janine Hacker}, \bibinfo{person}{Jan vom
  Brocke}, \bibinfo{person}{Joshua Handali}, \bibinfo{person}{Markus Otto},
  {and} \bibinfo{person}{Johannes Schneider}.} \bibinfo{year}{2020}\natexlab{}.
\newblock \showarticletitle{Virtually in this together--how web-conferencing
  systems enabled a new virtual togetherness during the COVID-19 crisis}.
\newblock \bibinfo{journal}{\emph{European Journal of Information Systems}}
  \bibinfo{volume}{29}, \bibinfo{number}{5} (\bibinfo{year}{2020}),
  \bibinfo{pages}{563--584}.
\newblock


\bibitem[He et~al\mbox{.}(2020)]%
        {he_preserving_2020}
\bibfield{author}{\bibinfo{person}{Yumei He}, \bibinfo{person}{Xingchen Xu},
  \bibinfo{person}{Ni Huang}, \bibinfo{person}{Yili Hong}, {and}
  \bibinfo{person}{De Liu}.} \bibinfo{year}{2020}\natexlab{}.
\newblock \bibinfo{booktitle}{\emph{Preserving {User} {Privacy} {Through}
  {Ephemeral} {Sharing} {Design}: {A} {Large}-{Scale} {Randomized} {Field}
  {Experiment} in the {Online} {Dating} {Context}}}.
\newblock \bibinfo{type}{{SSRN} {Scholarly} {Paper}} 3740782.
  \bibinfo{institution}{Social Science Research Network},
  \bibinfo{address}{Rochester, NY}.
\newblock
\urldef\tempurl%
\url{https://doi.org/10.2139/ssrn.3740782}
\showDOI{\tempurl}


\bibitem[Heilman and Parks-Stamm(2007)]%
        {heilman_gender_2007}
\bibfield{author}{\bibinfo{person}{Madeline~E. Heilman} {and}
  \bibinfo{person}{Elizabeth~J. Parks-Stamm}.} \bibinfo{year}{2007}\natexlab{}.
\newblock \showarticletitle{Gender {Stereotypes} in the {Workplace}:
  {Obstacles} to {Women}'s {Career} {Progress}}.
\newblock In \bibinfo{booktitle}{\emph{Social {Psychology} of {Gender}}},
  \bibfield{editor}{\bibinfo{person}{Shelley J.~Correll}} (Ed.).
  \bibinfo{series}{Advances in {Group} {Processes}}, Vol.~\bibinfo{volume}{24}.
  \bibinfo{publisher}{Emerald Group Publishing Limited},
  \bibinfo{pages}{47--77}.
\newblock
\showISBNx{978-0-7623-1430-0 978-1-84950-496-6}
\urldef\tempurl%
\url{https://doi.org/10.1016/S0882-6145(07)24003-2}
\showDOI{\tempurl}


\bibitem[Hughes and Dodge(1997)]%
        {hughes_african_1997}
\bibfield{author}{\bibinfo{person}{Diane Hughes} {and} \bibinfo{person}{Mark~A.
  Dodge}.} \bibinfo{year}{1997}\natexlab{}.
\newblock \showarticletitle{African {American} {Women} in the {Workplace}:
  {Relationships} {Between} {Job} {Conditions}, {Racial} {Bias} at {Work}, and
  {Perceived} {Job} {Quality}}.
\newblock \bibinfo{journal}{\emph{American Journal of Community Psychology}}
  \bibinfo{volume}{25}, \bibinfo{number}{5} (\bibinfo{year}{1997}),
  \bibinfo{pages}{581--599}.
\newblock
\showISSN{1573-2770}
\urldef\tempurl%
\url{https://doi.org/10.1023/A:1024630816168}
\showDOI{\tempurl}
\newblock
\shownote{\_eprint:
  https://onlinelibrary.wiley.com/doi/pdf/10.1023/A\%3A1024630816168}.


\bibitem[Jakobsson and Brock(2021)]%
        {jakobsson_energizing_2021}
\bibfield{author}{\bibinfo{person}{Ingrid~Bronken Jakobsson} {and}
  \bibinfo{person}{Tucker~Woodham Brock}.} \bibinfo{year}{2021}\natexlab{}.
\newblock \emph{\bibinfo{title}{Energizing the “Zoom-bie” Experience:
  Understanding virtual meetings through the influence of speaking times on
  perceived meeting satisfaction}}.
\newblock \bibinfo{thesistype}{Master's\ thesis}.
  \bibinfo{school}{Handelsh{\o}yskolen BI}.
\newblock


\bibitem[Kauffeld and Lehmann-Willenbrock(2012)]%
        {kauffeld_meetings_2012}
\bibfield{author}{\bibinfo{person}{Simone Kauffeld} {and} \bibinfo{person}{Nale
  Lehmann-Willenbrock}.} \bibinfo{year}{2012}\natexlab{}.
\newblock \showarticletitle{Meetings {Matter}: {Effects} of {Team} {Meetings}
  on {Team} and {Organizational} {Success}}.
\newblock \bibinfo{journal}{\emph{Small Group Research}} \bibinfo{volume}{43},
  \bibinfo{number}{2} (\bibinfo{date}{April} \bibinfo{year}{2012}),
  \bibinfo{pages}{130--158}.
\newblock
\showISSN{1046-4964}
\urldef\tempurl%
\url{https://doi.org/10.1177/1046496411429599}
\showDOI{\tempurl}
\newblock
\shownote{Publisher: SAGE Publications Inc}.


\bibitem[Kharroubi(2021)]%
        {kharroubi_global_2021}
\bibfield{author}{\bibinfo{person}{Daniela Kharroubi}.}
  \bibinfo{year}{2021}\natexlab{}.
\newblock \showarticletitle{Global {Workforce} {Diversity} {Management}:
  {Challenges} across the {World}}. In \bibinfo{booktitle}{\emph{{SHS} {Web} of
  {Conferences}}}, Vol.~\bibinfo{volume}{92}. \bibinfo{publisher}{EDP
  Sciences}, \bibinfo{address}{Les Ulis, France}.
\newblock
\urldef\tempurl%
\url{https://doi.org/10.1051/shsconf/20219202026}
\showDOI{\tempurl}
\newblock
\shownote{ISSN: 24165182 Section: Behavioral Economics and Decision-Making}.


\bibitem[Kreamer et~al\mbox{.}(2021)]%
        {kreamer_optimizing_2021}
\bibfield{author}{\bibinfo{person}{Liana Kreamer}, \bibinfo{person}{George
  Stock}, {and} \bibinfo{person}{Steven Rogelberg}.}
  \bibinfo{year}{2021}\natexlab{}.
\newblock \showarticletitle{Optimizing {Virtual} {Team} {Meetings}: {Attendee}
  and {Leader} {Perspectives}}.
\newblock \bibinfo{journal}{\emph{American Journal of Health Promotion}}
  \bibinfo{volume}{35}, \bibinfo{number}{5} (\bibinfo{date}{June}
  \bibinfo{year}{2021}), \bibinfo{pages}{744--747}.
\newblock
\showISSN{0890-1171}
\urldef\tempurl%
\url{https://doi.org/10.1177/08901171211007955e}
\showDOI{\tempurl}
\newblock
\shownote{Publisher: SAGE Publications Inc}.


\bibitem[Kydd and Ferry(1994)]%
        {Kydd_managerial_1994}
\bibfield{author}{\bibinfo{person}{Christine~T Kydd} {and}
  \bibinfo{person}{Diane~L Ferry}.} \bibinfo{year}{1994}\natexlab{}.
\newblock \showarticletitle{Managerial use of video conferencing}.
\newblock \bibinfo{journal}{\emph{Information \& Management}}
  \bibinfo{volume}{27}, \bibinfo{number}{6} (\bibinfo{year}{1994}),
  \bibinfo{pages}{369--375}.
\newblock


\bibitem[Lehmann-Willenbrock et~al\mbox{.}(2016)]%
        {lehmann-willenbrock_our_2016}
\bibfield{author}{\bibinfo{person}{N.K. Lehmann-Willenbrock},
  \bibinfo{person}{J.A. Allen}, {and} \bibinfo{person}{D. Belyeu}.}
  \bibinfo{year}{2016}\natexlab{}.
\newblock \showarticletitle{Our love/hate relationship with workplace meetings:
  {How} good and bad meeting attendee behaviors impact employee engagement and
  wellbeing}.
\newblock \bibinfo{journal}{\emph{Management Research Review}}
  \bibinfo{volume}{39} (\bibinfo{year}{2016}), \bibinfo{pages}{1293--1312}.
\newblock
\showISSN{2040-8277}
\urldef\tempurl%
\url{https://doi.org/10.1108/MRR-08-2015-0195}
\showDOI{\tempurl}


\bibitem[Lehmann-Willenbrock et~al\mbox{.}(2017)]%
        {lehmann_critical_2017}
\bibfield{author}{\bibinfo{person}{Nale Lehmann-Willenbrock},
  \bibinfo{person}{Steven~G Rogelberg}, \bibinfo{person}{Joseph~A Allen}, {and}
  \bibinfo{person}{John~E Kello}.} \bibinfo{year}{2017}\natexlab{}.
\newblock \showarticletitle{The critical importance of meetings to leader and
  organizational success: Evidence-based insights and implications for key
  stakeholders}.
\newblock \bibinfo{journal}{\emph{Organizational Dynamics}}
  \bibinfo{volume}{47}, \bibinfo{number}{1} (\bibinfo{year}{2017}),
  \bibinfo{pages}{32}.
\newblock


\bibitem[Lekchiri et~al\mbox{.}(2019)]%
        {lekchiri_perceived_2019}
\bibfield{author}{\bibinfo{person}{Siham Lekchiri}, \bibinfo{person}{Cindy
  Crowder}, \bibinfo{person}{Anna Schnerre}, {and}
  \bibinfo{person}{Barbara~A.W. Eversole}.} \bibinfo{year}{2019}\natexlab{}.
\newblock \showarticletitle{Perceived workplace gender-bias and psychological
  impact: {The} case of women in a {Moroccan} higher education institution}.
\newblock \bibinfo{journal}{\emph{European Journal of Training and
  Development}} \bibinfo{volume}{43}, \bibinfo{number}{3/4}
  (\bibinfo{date}{Jan.} \bibinfo{year}{2019}), \bibinfo{pages}{339--353}.
\newblock
\showISSN{2046-9012}
\urldef\tempurl%
\url{https://doi.org/10.1108/EJTD-09-2018-0088}
\showDOI{\tempurl}
\newblock
\shownote{Publisher: Emerald Publishing Limited}.


\bibitem[Levanon(2020)]%
        {levanon_remote_2020}
\bibfield{author}{\bibinfo{person}{Gad Levanon}.}
  \bibinfo{year}{2020}\natexlab{}.
\newblock \bibinfo{title}{Remote {Work}: {The} {Biggest} {Legacy} {Of}
  {Covid}-19}.
\newblock
\newblock
\urldef\tempurl%
\url{https://www.forbes.com/sites/gadlevanon/2020/11/23/remote-work-the-biggest-legacy-of-covid-19/}
\showURL{%
\tempurl}
\newblock
\shownote{Section: Manufacturing}.


\bibitem[Long and Van~Dam(2019)]%
        {long_women_2019}
\bibfield{author}{\bibinfo{person}{Heather Long} {and} \bibinfo{person}{Andrew
  Van~Dam}.} \bibinfo{year}{2019}\natexlab{}.
\newblock \bibinfo{title}{Women of color are surging into the {U}.{S}.
  workforce, causing a historic first in who's getting hired - {The}
  {Washington} {Post}}.
\newblock
\newblock
\urldef\tempurl%
\url{https://www.washingtonpost.com/business/economy/for-the-first-time-ever-most-new-working-age-hires-in-the-us-are-people-of-color/2019/09/09/8edc48a2-bd10-11e9-b873-63ace636af08_story.html}
\showURL{%
\tempurl}


\bibitem[Ma et~al\mbox{.}(2021)]%
        {ma_perceived_2021}
\bibfield{author}{\bibinfo{person}{Xiaofen Ma}, \bibinfo{person}{Yuren Qin},
  \bibinfo{person}{Zhuo Chen}, {and} \bibinfo{person}{Hichang Cho}.}
  \bibinfo{year}{2021}\natexlab{}.
\newblock \showarticletitle{Perceived ephemerality, privacy calculus, and the
  privacy settings of an ephemeral social media site}.
\newblock \bibinfo{journal}{\emph{Computers in Human Behavior}}
  \bibinfo{volume}{124} (\bibinfo{date}{Nov.} \bibinfo{year}{2021}),
  \bibinfo{pages}{106928}.
\newblock
\showISSN{0747-5632}
\urldef\tempurl%
\url{https://doi.org/10.1016/j.chb.2021.106928}
\showDOI{\tempurl}


\bibitem[Militello(2015)]%
        {militello_managing_2015}
\bibfield{author}{\bibinfo{person}{Kyle Militello}.}
  \bibinfo{year}{2015}\natexlab{}.
\newblock \showarticletitle{Managing a {Homosexual} {Identity} within a
  {Heteronormative} {Workplace} {Environment}}.
\newblock \bibinfo{journal}{\emph{The Sociological Imagination: Undergraduate
  Journal}} \bibinfo{volume}{4}, \bibinfo{number}{1} (\bibinfo{date}{July}
  \bibinfo{year}{2015}).
\newblock
\urldef\tempurl%
\url{https://ojs.lib.uwo.ca/index.php/si/article/view/5295}
\showURL{%
\tempurl}
\newblock
\shownote{Number: 1}.


\bibitem[Moceri(2012)]%
        {moceri_bias_2012}
\bibfield{author}{\bibinfo{person}{Joane~T. Moceri}.}
  \bibinfo{year}{2012}\natexlab{}.
\newblock \showarticletitle{Bias in the {Nursing} {Workplace}: {Implications}
  for {Latino}(a) {Nurses}}.
\newblock \bibinfo{journal}{\emph{Journal of Cultural Diversity}}
  \bibinfo{volume}{19}, \bibinfo{number}{3} (\bibinfo{year}{2012}),
  \bibinfo{pages}{94--101}.
\newblock
\showISSN{10715568}
\urldef\tempurl%
\url{https://search.ebscohost.com/login.aspx?direct=true&db=aph&AN=79968254&site=eds-live}
\showURL{%
\tempurl}
\newblock
\shownote{Publisher: Tucker Publications, Inc.}.


\bibitem[Nguyen and Canny(2007)]%
        {nguyen_multiview_2007}
\bibfield{author}{\bibinfo{person}{David~T Nguyen} {and} \bibinfo{person}{John
  Canny}.} \bibinfo{year}{2007}\natexlab{}.
\newblock \showarticletitle{Multiview: improving trust in group video
  conferencing through spatial faithfulness}. In
  \bibinfo{booktitle}{\emph{Proceedings of the SIGCHI conference on Human
  factors in computing systems}}. \bibinfo{pages}{1465--1474}.
\newblock


\bibitem[Noviyanti(2022)]%
        {noviyanti_students_2022}
\bibfield{author}{\bibinfo{person}{Dinda~Rachmitha Noviyanti}.}
  \bibinfo{year}{2022}\natexlab{}.
\newblock \showarticletitle{{STUDENTS}’ {PERCEPTION} {ON} {THE} {UTILIZATION}
  {OF} {ZOOM} {VIDEO} {CONFERENCING} {IN} {IMPROVING} {SPEAKING} {ABILITY}
  {AMIDST} {COVID}-19 {PANDEMIC}}.
\newblock \bibinfo{journal}{\emph{ELLTER Journal}} \bibinfo{volume}{3},
  \bibinfo{number}{1} (\bibinfo{date}{April} \bibinfo{year}{2022}),
  \bibinfo{pages}{1--8}.
\newblock
\showISSN{2746-1424}
\urldef\tempurl%
\url{https://doi.org/10.22236/ellter.v3i1.8956}
\showDOI{\tempurl}
\newblock
\shownote{Number: 1 Publisher: Universitas Muhammadiyah Prof. DR. Hamka (UHAMKA
  Press)}.


\bibitem[Nowrouzi et~al\mbox{.}(2015)]%
        {nowrouzi_weight_2015}
\bibfield{author}{\bibinfo{person}{Behdin Nowrouzi}, \bibinfo{person}{Alicia
  McDougall}, \bibinfo{person}{Basem Gohar}, \bibinfo{person}{Behnam
  Nowrouz-Kia}, \bibinfo{person}{Jennifer Casole}, {and} \bibinfo{person}{Fizza
  Ali}.} \bibinfo{year}{2015}\natexlab{}.
\newblock \showarticletitle{Weight bias in the workplace: A literature review}.
\newblock \bibinfo{journal}{\emph{Occupational Medicine \& Health Affairs}}
  (\bibinfo{year}{2015}).
\newblock


\bibitem[Nunneley et~al\mbox{.}(2021)]%
        {nunneley_leading_2021}
\bibfield{author}{\bibinfo{person}{Chloë~E. Nunneley},
  \bibinfo{person}{Michael Fishman}, \bibinfo{person}{Kathryn~M. Sundheim},
  \bibinfo{person}{Rachel~E. Korus}, \bibinfo{person}{Robert~H. Rosen},
  \bibinfo{person}{Blair~A. Streater}, \bibinfo{person}{Katherine~A.
  O’Donnell}, \bibinfo{person}{Lori~R. Newman}, {and}
  \bibinfo{person}{Carolyn~H. Marcus}.} \bibinfo{year}{2021}\natexlab{}.
\newblock \showarticletitle{Leading synchronous virtual teaching sessions}.
\newblock \bibinfo{journal}{\emph{The Clinical Teacher}} \bibinfo{volume}{18},
  \bibinfo{number}{3} (\bibinfo{year}{2021}), \bibinfo{pages}{231--235}.
\newblock
\showISSN{1743-498X}
\urldef\tempurl%
\url{https://doi.org/10.1111/tct.13282}
\showDOI{\tempurl}
\newblock
\shownote{\_eprint: https://onlinelibrary.wiley.com/doi/pdf/10.1111/tct.13282}.


\bibitem[Okereafor and Manny(2020)]%
        {okereafor_understanding_2020}
\bibfield{author}{\bibinfo{person}{Kenneth Okereafor} {and}
  \bibinfo{person}{Phil Manny}.} \bibinfo{year}{2020}\natexlab{}.
\newblock \showarticletitle{Understanding cybersecurity challenges of
  telecommuting and video conferencing applications in the COVID-19 pandemic}.
\newblock \bibinfo{journal}{\emph{Journal Homepage: http://ijmr. net. in}}
  \bibinfo{volume}{8}, \bibinfo{number}{6} (\bibinfo{year}{2020}).
\newblock


\bibitem[Olson and Olson(2000)]%
        {olson_distance_2000}
\bibfield{author}{\bibinfo{person}{Gary~M Olson} {and}
  \bibinfo{person}{Judith~S Olson}.} \bibinfo{year}{2000}\natexlab{}.
\newblock \showarticletitle{Distance matters}.
\newblock \bibinfo{journal}{\emph{Human--computer interaction}}
  \bibinfo{volume}{15}, \bibinfo{number}{2-3} (\bibinfo{year}{2000}),
  \bibinfo{pages}{139--178}.
\newblock


\bibitem[Perkins(2009)]%
        {perkins_executive_2009}
\bibfield{author}{\bibinfo{person}{Robert~D Perkins}.}
  \bibinfo{year}{2009}\natexlab{}.
\newblock \showarticletitle{How executive coaching can change leader behavior
  and improve meeting effectiveness: An exploratory study.}
\newblock \bibinfo{journal}{\emph{Consulting Psychology Journal: Practice and
  Research}} \bibinfo{volume}{61}, \bibinfo{number}{4} (\bibinfo{year}{2009}),
  \bibinfo{pages}{298}.
\newblock


\bibitem[Ricard(2020)]%
        {ricard_council_2020}
\bibfield{author}{\bibinfo{person}{Sébastien Ricard}.}
  \bibinfo{year}{2020}\natexlab{}.
\newblock \bibinfo{title}{Council {Post}: {The} {Year} {Of} {The} {Knowledge}
  {Worker}}.
\newblock
\newblock
\urldef\tempurl%
\url{https://www.forbes.com/sites/forbestechcouncil/2020/12/10/the-year-of-the-knowledge-worker/}
\showURL{%
\tempurl}
\newblock
\shownote{Section: Innovation}.


\bibitem[Richter(2021)]%
        {richter_infographic_2021}
\bibfield{author}{\bibinfo{person}{Felix Richter}.}
  \bibinfo{year}{2021}\natexlab{}.
\newblock \bibinfo{title}{Infographic: {Zoom} {Retains} {Pandemic} {Gains} {As}
  {Hybrid} {Work} {Is} {Here} to {Stay}}.
\newblock
\newblock
\urldef\tempurl%
\url{https://www.statista.com/chart/21906/zoom-revenue/}
\showURL{%
\tempurl}


\bibitem[Romano and Nunamaker(2001)]%
        {romano_meeting_2001}
\bibfield{author}{\bibinfo{person}{N.C. Romano} {and} \bibinfo{person}{J.F.
  Nunamaker}.} \bibinfo{year}{2001}\natexlab{}.
\newblock \showarticletitle{Meeting analysis: findings from research and
  practice}. In \bibinfo{booktitle}{\emph{Proceedings of the 34th {Annual}
  {Hawaii} {International} {Conference} on {System} {Sciences}}}.
  \bibinfo{pages}{13 pp.--}.
\newblock
\urldef\tempurl%
\url{https://doi.org/10.1109/HICSS.2001.926253}
\showDOI{\tempurl}


\bibitem[Rommelse et~al\mbox{.}(2020)]%
        {rommelse_slow_2020}
\bibfield{author}{\bibinfo{person}{Nanda Rommelse}, \bibinfo{person}{Marjolein
  Luman}, {and} \bibinfo{person}{Rogier Kievit}.}
  \bibinfo{year}{2020}\natexlab{}.
\newblock \showarticletitle{Slow processing speed: a cross-disorder phenomenon
  with significant clinical value, and in need of further methodological
  scrutiny}.
\newblock \bibinfo{journal}{\emph{European Child \& Adolescent Psychiatry}}
  \bibinfo{volume}{29}, \bibinfo{number}{10} (\bibinfo{date}{Oct.}
  \bibinfo{year}{2020}), \bibinfo{pages}{1325--1327}.
\newblock
\showISSN{1435-165X}
\urldef\tempurl%
\url{https://doi.org/10.1007/s00787-020-01639-9}
\showDOI{\tempurl}


\bibitem[Rudolph et~al\mbox{.}(2009)]%
        {rudolph_meta_2009}
\bibfield{author}{\bibinfo{person}{Cort~W Rudolph}, \bibinfo{person}{Charles~L
  Wells}, \bibinfo{person}{Marcus~D Weller}, {and} \bibinfo{person}{Boris~B
  Baltes}.} \bibinfo{year}{2009}\natexlab{}.
\newblock \showarticletitle{A meta-analysis of empirical studies of
  weight-based bias in the workplace}.
\newblock \bibinfo{journal}{\emph{Journal of Vocational Behavior}}
  \bibinfo{volume}{74}, \bibinfo{number}{1} (\bibinfo{year}{2009}),
  \bibinfo{pages}{1--10}.
\newblock


\bibitem[Satar and Özdener(2008)]%
        {satar_effects_2008}
\bibfield{author}{\bibinfo{person}{H.~Müge Satar} {and}
  \bibinfo{person}{Nesrin Özdener}.} \bibinfo{year}{2008}\natexlab{}.
\newblock \showarticletitle{The {Effects} of {Synchronous} {CMC} on {Speaking}
  {Proficiency} and {Anxiety}: {Text} {Versus} {Voice} {Chat}}.
\newblock \bibinfo{journal}{\emph{The Modern Language Journal}}
  \bibinfo{volume}{92}, \bibinfo{number}{4} (\bibinfo{year}{2008}),
  \bibinfo{pages}{595--613}.
\newblock
\showISSN{1540-4781}
\urldef\tempurl%
\url{https://doi.org/10.1111/j.1540-4781.2008.00789.x}
\showDOI{\tempurl}
\newblock
\shownote{\_eprint:
  https://onlinelibrary.wiley.com/doi/pdf/10.1111/j.1540-4781.2008.00789.x}.


\bibitem[Sauer and Kauffeld(2015)]%
        {sauer_ties_2015}
\bibfield{author}{\bibinfo{person}{Nils~Christian Sauer} {and}
  \bibinfo{person}{Simone Kauffeld}.} \bibinfo{year}{2015}\natexlab{}.
\newblock \showarticletitle{The ties of meeting leaders: A social network
  analysis}.
\newblock \bibinfo{journal}{\emph{Psychology}} \bibinfo{volume}{6},
  \bibinfo{number}{04} (\bibinfo{year}{2015}), \bibinfo{pages}{415}.
\newblock


\bibitem[Schieman et~al\mbox{.}(2021)]%
        {schieman_work-life_2021}
\bibfield{author}{\bibinfo{person}{Scott Schieman}, \bibinfo{person}{Philip~J.
  Badawy}, \bibinfo{person}{Melissa A.~Milkie}, {and} \bibinfo{person}{Alex
  Bierman}.} \bibinfo{year}{2021}\natexlab{}.
\newblock \showarticletitle{Work-{Life} {Conflict} {During} the {COVID}-19
  {Pandemic}}.
\newblock \bibinfo{journal}{\emph{Socius}}  \bibinfo{volume}{7}
  (\bibinfo{date}{Jan.} \bibinfo{year}{2021}),
  \bibinfo{pages}{2378023120982856}.
\newblock
\showISSN{2378-0231}
\urldef\tempurl%
\url{https://doi.org/10.1177/2378023120982856}
\showDOI{\tempurl}
\newblock
\shownote{Publisher: SAGE Publications}.


\bibitem[Scholl et~al\mbox{.}(2006)]%
        {scholl_comparison_2006}
\bibfield{author}{\bibinfo{person}{Jeremiah Scholl}, \bibinfo{person}{John
  McCarthy}, {and} \bibinfo{person}{Rikard Harr}.}
  \bibinfo{year}{2006}\natexlab{}.
\newblock \showarticletitle{A comparison of chat and audio in media rich
  environments}. In \bibinfo{booktitle}{\emph{Proceedings of the 2006 20th
  anniversary conference on {Computer} supported cooperative work}}
  \emph{(\bibinfo{series}{{CSCW} '06})}. \bibinfo{publisher}{Association for
  Computing Machinery}, \bibinfo{address}{New York, NY, USA},
  \bibinfo{pages}{323--332}.
\newblock
\showISBNx{978-1-59593-249-5}
\urldef\tempurl%
\url{https://doi.org/10.1145/1180875.1180925}
\showDOI{\tempurl}


\bibitem[Schuleigh et~al\mbox{.}(2019)]%
        {schuleigh_enhancing_2019}
\bibfield{author}{\bibinfo{person}{Vivien~E Schuleigh}, \bibinfo{person}{John~M
  Malouff}, \bibinfo{person}{Nicola~S Schutte}, {and}
  \bibinfo{person}{Natasha~M Loi}.} \bibinfo{year}{2019}\natexlab{}.
\newblock \showarticletitle{ENHANCING MEETINGS: The Impact of Leader Behavior.}
\newblock \bibinfo{journal}{\emph{Journal of Leadership Education}}
  \bibinfo{volume}{18}, \bibinfo{number}{3} (\bibinfo{year}{2019}).
\newblock


\bibitem[Schuleigh et~al\mbox{.}(2021)]%
        {schuleigh_effects_2021}
\bibfield{author}{\bibinfo{person}{Vivien~E Schuleigh}, \bibinfo{person}{John~M
  Malouff}, \bibinfo{person}{Nicola~S Schutte}, {and}
  \bibinfo{person}{Natasha~M Loi}.} \bibinfo{year}{2021}\natexlab{}.
\newblock \showarticletitle{Effects of Meeting Leader Training on Meeting
  Attendees.}
\newblock \bibinfo{journal}{\emph{Journal of Leadership Education}}
  \bibinfo{volume}{20}, \bibinfo{number}{1} (\bibinfo{year}{2021}).
\newblock


\bibitem[Scott et~al\mbox{.}(2015)]%
        {scott_five_2015}
\bibfield{author}{\bibinfo{person}{Cliff Scott}, \bibinfo{person}{Joseph~A.
  Allen}, \bibinfo{person}{Steven~G. Rogelberg}, {and} \bibinfo{person}{Alex
  Kello}.} \bibinfo{year}{2015}\natexlab{}.
\newblock \showarticletitle{Five {Theoretical} {Lenses} for {Conceptualizing}
  the {Role} of {Meetings} in {Organizational} {Life}}.
\newblock In \bibinfo{booktitle}{\emph{The {Cambridge} {Handbook} of {Meeting}
  {Science}}}. \bibinfo{publisher}{Cambridge University Press},
  \bibinfo{address}{Cambridge}, \bibinfo{pages}{20--46}.
\newblock
\showISBNx{978-1-107-06718-9}
\urldef\tempurl%
\url{https://doi.org/10.1017/CBO9781107589735.003}
\showDOI{\tempurl}


\bibitem[Smolenskiy and Levshin(2022)]%
        {smolenskiy_problem_2022}
\bibfield{author}{\bibinfo{person}{Mikhail Smolenskiy} {and}
  \bibinfo{person}{Nikolay Levshin}.} \bibinfo{year}{2022}\natexlab{}.
\newblock \showarticletitle{The {Problem} of {Determining} the {Video}
  {Conferencing} {Platform} {Criteria} for {Online} {Learning}}. In
  \bibinfo{booktitle}{\emph{{XIV} {International} {Scientific} {Conference}
  “{INTERAGROMASH} 2021”}} \emph{(\bibinfo{series}{Lecture {Notes} in
  {Networks} and {Systems}})}, \bibfield{editor}{\bibinfo{person}{Alexey
  Beskopylny} {and} \bibinfo{person}{Mark Shamtsyan}} (Eds.).
  \bibinfo{publisher}{Springer International Publishing},
  \bibinfo{address}{Cham}, \bibinfo{pages}{357--364}.
\newblock
\showISBNx{978-3-030-80946-1}
\urldef\tempurl%
\url{https://doi.org/10.1007/978-3-030-80946-1_35}
\showDOI{\tempurl}


\bibitem[Steinzor(1950)]%
        {steinzor_spatial_1950}
\bibfield{author}{\bibinfo{person}{Bernard Steinzor}.}
  \bibinfo{year}{1950}\natexlab{}.
\newblock \showarticletitle{The spatial factor in face to face discussion
  groups}.
\newblock \bibinfo{journal}{\emph{The Journal of Abnormal and Social
  Psychology}} \bibinfo{volume}{45}, \bibinfo{number}{3}
  (\bibinfo{year}{1950}), \bibinfo{pages}{552--555}.
\newblock
\showISSN{0096-851X}
\urldef\tempurl%
\url{https://doi.org/10.1037/h0061767}
\showDOI{\tempurl}
\newblock
\shownote{Place: US Publisher: American Psychological Association}.


\bibitem[Stoodley and Stein(2006)]%
        {stoodley_processing_2006}
\bibfield{author}{\bibinfo{person}{Catherine~J. Stoodley} {and}
  \bibinfo{person}{John~F. Stein}.} \bibinfo{year}{2006}\natexlab{}.
\newblock \showarticletitle{A processing speed deficit in dyslexic adults?
  {Evidence} from a peg-moving task}.
\newblock \bibinfo{journal}{\emph{Neuroscience Letters}} \bibinfo{volume}{399},
  \bibinfo{number}{3} (\bibinfo{date}{May} \bibinfo{year}{2006}),
  \bibinfo{pages}{264--267}.
\newblock
\showISSN{0304-3940}
\urldef\tempurl%
\url{https://doi.org/10.1016/j.neulet.2006.02.004}
\showDOI{\tempurl}


\bibitem[Stoodley and Stein(2011)]%
        {stoodley_cerebellum_2011}
\bibfield{author}{\bibinfo{person}{Catherine~J. Stoodley} {and}
  \bibinfo{person}{John~F. Stein}.} \bibinfo{year}{2011}\natexlab{}.
\newblock \showarticletitle{The cerebellum and dyslexia}.
\newblock \bibinfo{journal}{\emph{Cortex: A Journal Devoted to the Study of the
  Nervous System and Behavior}} \bibinfo{volume}{47}, \bibinfo{number}{1}
  (\bibinfo{year}{2011}), \bibinfo{pages}{101--116}.
\newblock
\showISSN{1973-8102}
\urldef\tempurl%
\url{https://doi.org/10.1016/j.cortex.2009.10.005}
\showDOI{\tempurl}
\newblock
\shownote{Place: France Publisher: Elsevier Masson SAS}.


\bibitem[Thompson et~al\mbox{.}(2009)]%
        {thompson_supporting_2009}
\bibfield{author}{\bibinfo{person}{Phil Thompson}, \bibinfo{person}{Rahat
  Iqbal}, {and} \bibinfo{person}{Anne James}.} \bibinfo{year}{2009}\natexlab{}.
\newblock \showarticletitle{Supporting collaborative virtual meetings using
  multi-agent systems}. In \bibinfo{booktitle}{\emph{2009 13th International
  Conference on Computer Supported Cooperative Work in Design}}. IEEE,
  \bibinfo{pages}{276--281}.
\newblock


\bibitem[Toossi(2015)]%
        {toossi_labor_2015}
\bibfield{author}{\bibinfo{person}{Mitra Toossi}.}
  \bibinfo{year}{2015}\natexlab{}.
\newblock \bibinfo{title}{Labor force projections to 2024: the labor force is
  growing, but slowly : {Monthly} {Labor} {Review}: {U}.{S}. {Bureau} of
  {Labor} {Statistics}}.
\newblock
\newblock
\urldef\tempurl%
\url{https://www.bls.gov/opub/mlr/2015/article/labor-force-projections-to-2024.htm}
\showURL{%
\tempurl}


\bibitem[Williams(2021)]%
        {williams_working_2021}
\bibfield{author}{\bibinfo{person}{Nerys Williams}.}
  \bibinfo{year}{2021}\natexlab{}.
\newblock \showarticletitle{Working through COVID-19:‘Zoom’gloom and
  ‘Zoom’fatigue}.
\newblock \bibinfo{journal}{\emph{Occupational Medicine}} \bibinfo{volume}{71},
  \bibinfo{number}{3} (\bibinfo{year}{2021}), \bibinfo{pages}{164--164}.
\newblock


\bibitem[Zhu et~al\mbox{.}(2015)]%
        {zhu_employee_2015}
\bibfield{author}{\bibinfo{person}{Ying Zhu}, \bibinfo{person}{Yuhua Xie},
  \bibinfo{person}{Malcolm Warner}, {and} \bibinfo{person}{Yongxing Guo}.}
  \bibinfo{year}{2015}\natexlab{}.
\newblock \showarticletitle{Employee participation and the influence on job
  satisfaction of the ‘new generation’ of {Chinese} employees}.
\newblock \bibinfo{journal}{\emph{The International Journal of Human Resource
  Management}} \bibinfo{volume}{26}, \bibinfo{number}{19} (\bibinfo{date}{Oct.}
  \bibinfo{year}{2015}), \bibinfo{pages}{2395--2411}.
\newblock
\showISSN{0958-5192}
\urldef\tempurl%
\url{https://doi.org/10.1080/09585192.2014.990397}
\showDOI{\tempurl}
\newblock
\shownote{Publisher: Routledge \_eprint:
  https://doi.org/10.1080/09585192.2014.990397}.


\end{thebibliography}
\end{document}